\documentclass[fleqn,usenatbib]{mnras}

\usepackage{newtxtext,newtxmath}

\usepackage[T1]{fontenc}
\usepackage{ulem}

\DeclareRobustCommand{\VAN}[3]{#2}
\let\VANthebibliography\thebibliography
\def\thebibliography{\DeclareRobustCommand{\VAN}[3]{##3}\VANthebibliography}

\usepackage{graphicx}	
\usepackage{amsmath}	
\usepackage{soul}       
\usepackage{pifont}
\usepackage{fontawesome5}
\usepackage{hyperref}

\newcommand{\todo}[1]{\textcolor{black}{#1}}

\newcommand{\jc}[1]{\textcolor{cyan}{#1}}
\newcommand{\xmark}{\ding{55}}

\title[Modeling the Astrometric Performance of MAVIS]{Towards Realistic Modeling of the Astrometric Capabilities of MCAO Systems: Detecting an \todo{Intermediate Mass Black Hole} with MAVIS}

\author[S. Monty et al.]{Stephanie Monty$^{1,2}$\thanks{E-mail: Stephanie.Monty@anu.edu.au},
Fran\c{c}ois Rigaut$^{1}$,
Richard McDermid$^{3,2}$,
Holger Baumgardt$^{4}$,
\newauthor 
Jesse Cranney$^{1}$,
Guido Agapito$^{5}$,
J. Trevor Mendel$^{1,2}$,
C\'{e}dric Plantet$^{5}$,
\newauthor 
Davide Greggio$^{6}$,
Peter B. Stetson$^{7}$,
Giuliana Fiorentino$^{8}$,
Dionne Haynes$^{1}$
\\
$^{1}$ Research School of Astronomy and Astrophysics, Mt Stromlo Observatory, Weston Creek, ACT 2611, Australia\\
$^{2}$ ARC Centre of Excellence for All Sky Astrophysics in 3 Dimensions (ASTRO 3D), Australia\\
$^{3}$ Research Centre for Astronomy, Astrophysics, and Astrophotonics, Department of Physics and Astronomy, Macquarie University, NSW 2109, Australia\\
$^{4}$ School of Mathematics and Physics, The University of Queensland, St.Lucia, QLD 4072, Australia\\
$^{5}$ INAF – Osservatorio Astronomico di Arcetri, Largo Enrico Fermi 5, I-50125 Firenze, Italy\\
$^{6}$ INAF - Osservatorio Astronomico di Padova, Vicolo dell’Osservatorio 5, 35122 Padova, Italy\\
$^{7}$ Herzberg Astronomy and Astrophysics, National Research Council, 5071 West Saanich Road, Victoria, British Columbia V9E 2E7, Canada\\
$^{8}$ INAF – Osservatorio Astronomico di Roma, via Frascati 33, I-00078 Monte Porzio Catone, Roma, Italy
}

\date{Accepted XXX. Received YYY; in original form ZZZ}

\pubyear{2020}

\begin{document}
\label{firstpage}
\pagerange{\pageref{firstpage}--\pageref{lastpage}}
\maketitle

\begin{abstract}
Accurate astrometry is a key deliverable for the next generation of multi-conjugate adaptive optics (MCAO) systems. The MCAO Visible Imager and Spectrograph (MAVIS) is being designed for the Very Large Telescope Adaptive Optics Facility and must achieve 150~$\mu$as astrometric precision (50~$\mu$as goal). To test this before going on-sky, we have created \texttt{MAVISIM}, a tool to simulate MAVIS images. \texttt{MAVISIM} accounts for three major sources of astrometric error, high- and low-order point spread function (PSF) spatial variability, tip-tilt residual error and static field distortion. \todo{When exploring the impact of these three error terms alone}, we recover an astrometric accuracy of 50~$\mu$as for all stars brighter than $m=19$ in a 30s integration using PSF-fitting photometry. We also assess the feasibility of MAVIS detecting an intermediate mass black hole (IMBH) in a Milky Way globular cluster. We use an N-body simulation of an NGC~3201-like cluster with a central 1500 M$_{\odot}$ IMBH as input to \texttt{MAVISIM} and recover the velocity dispersion profile from proper motion measurements. \todo{Under favourable astrometric conditions}, the dynamical signature of the IMBH is detected with a precision of $\sim0.20$~km/s in the inner $\sim4\arcsec$ of the cluster where HST is confusion-limited. This precision is comparable to measurements made by \textit{Gaia}, HST and MUSE in the outer $\sim60\arcsec$ of the cluster. This study is the first step towards building a science-driven astrometric error budget for an MCAO system and a prediction of what MAVIS could do once on sky. \href{https://github.com/smonty93/MAVISIM}{\faGithub}
\end{abstract}

\begin{keywords}
instrumentation: adaptive optics -- astrometry -- proper motions -- stars: black holes -- (Galaxy:) globular clusters: general -- (Galaxy:) globular clusters: individual: NGC 3201
\end{keywords}



\section{\label{sec:intro} Introduction}
Precision astrometry is entering a new era, ushered in by the third (early) data release of the \textit{Gaia} satellite \citep{gaiadr1, gaiadr2, gaiadr3} and surely culminating in the first measurements from the European Extremely Large Telescope's MICADO imager \citep[Multi-AO Imaging Camera for Deep Observations,][]{micado}. While \textit{Gaia} has provided exceptional astrometry of single stellar sources out to large distances, its accuracy is inherently limited in crowded fields, especially those beyond the Milky Way (MW) Halo. This regime will be reserved for the extremely large telescopes (ELTs). As an example, the centre of Milky Way (MW) globular clusters (GCs) are regions complicated by both crowding and often distance. The Hubble Space Telescope (HST) has dominated the study of crowded cluster centres since its inception, making use of its diffraction-limited imaging and near-distortion-free reference frame \citep{anderson03} to produce a plethora of surveys and dedicated studies of MW GCs \citep{sarajedini07, milone09, piotto15, milone17}. HST has performed especially well in the field of precision proper motions (PMs), where observations over multiple epochs have revealed the often complex internal kinematics of globular clusters \citep{anderson98, mclaughlin06, hstpromo1, massari2015, hstpromo2, bellini17, libralato18, raso2020}. 

Competing with HST from the ground has only been possible in the last couple of decades, with the implementation of adaptive optics (AO) systems on 8m-class telescopes \citep{altair, vltao, keckao, subaruao, lbtao}. See \cite{rigaut2018} and references therein for a review of the field. Given the projected on-sky size of MW GCs ($\sim60''$), they have been the quintessential objects to study. Multi-conjugate adaptive optics (MCAO), a flavour of wide field AO, provides the most uniform AO correction across a wide field of view (FoV) by correcting multiple layers of the atmosphere. This is done using multiple deformable mirrors and natural and laser guide stars \citep[NGS and LGS respectively,][]{dicke1975, beckers1988, ellerbroek1994}. Since the first demonstration of MCAO on-sky in 2007 using ESO's MCAO Demonstrator MAD, astrometry with MCAO has become increasingly popular \citep{madmcao, madonsky, madastrometry}. The only MCAO system on-sky today (used for night time astronomy) is the Gemini Multi-conjugate adaptive optics System \citep[GeMS,][]{gems1, gems2} at the Gemini South Telescope. Although GeMS was not designed with accurate astrometry in mind, it has since been shown to be a powerful tool for both astrometric and PM measurements in MW GCs \citep{massari2015, massari16b, dalessandro16, monty2018}.

\begin{figure}
\centering
\includegraphics[scale=0.35]{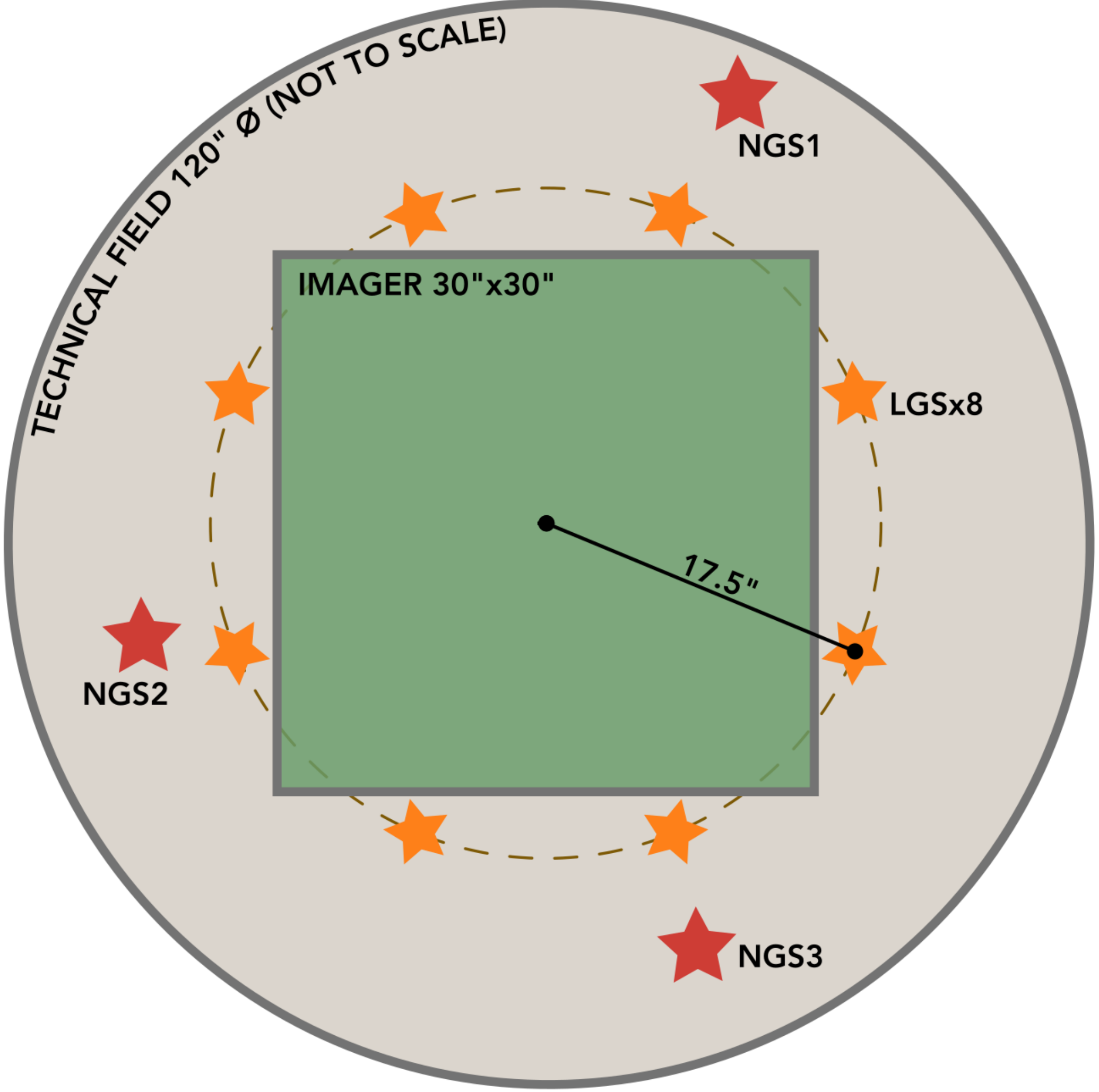}
\caption{\label{fig:mavisschem} Schematic of the MAVIS on-sky footprint showing; the technical field over which the NGS (shown in red) will be selected ($r_{\mathrm{PF}}=60\arcsec$), the projected size of the CCD (shown as the green square) covering the science field ($r_{\mathrm{SF}}=15\arcsec$), and the eight LGS (shown in orange).
}
\end{figure}

As part of the next generation of MCAO instruments, the MCAO Assisted Visible Imager and Spectrograph (MAVIS) is an instrument being built for the Very Large Telescope Adaptive Optics Facility \citep{rigautmavis, mavisscience}. MAVIS will be the first workhorse MCAO system to perform AO corrections in the visible,  with performance optimised at 550~nm. To do this, MAVIS will use three deformable mirrors conjugated to correct the ground layer, 6~km and 13.5~km layer turbulence, eight LGS and three NGS. For clarity, a schematic of the MAVIS on-sky footprint is shown in Fig.~\ref{fig:mavisschem}, where the circular technical field is marked ($r_{\mathrm{TF}}=60\arcsec$), the CCD footprint is shown as the green region enclosing the AO-corrected science field ($r_{\mathrm{SF}}=15\arcsec$). The three NGS (shown in red in Fig.\ref{fig:mavisschem}) are selected within the technical field, while the eight LGS (shown in orange) lie just beyond the science field. MAVIS will provide 15\% encircled energy within 50~mas over 50\% of the sky at the South Galactic Pole. Furthermore, MAVIS will deliver a maximum FWHM of 26~mas (goal 24~mas) using three NGS as faint as magnitude $H=18$. 

Feeding both an imager and a spectrograph, a core science target of MAVIS will be crowded stellar fields. A science target shared in common with MICADO \citep{micadoscience}. To complement the capabilities of MICADO, MAVIS will add additional spectral information below 800~nm \textit{and} deliver astrometric and PM accuracy on the order equal to the ELT with a \textit{differential} astrometric requirement of $150\jc{\ }\mu$as and target of $50\jc{\ }\mu$as. To achieve this level of astrometric accuracy, each term contributing to the final astrometric error must be well understood and accounted for in the final astrometric error budget. This is challenging, as MCAO systems have been shown to have both static and dynamic astrometric error terms \citep{lu14, massari16b, patti2019}, with dynamic contributions varying on the timescale of single exposures. Despite the challenges associated with astrometry on MCAO instruments, they provide the only opportunity to achieve similar or better spatial resolution than HST in dense stellar fields over a comparable FoV \citep{dalessandro16, massari16b}.

MAVIS will explore a variety of science cases in crowded stellar fields, with a key science theme being the study of stellar clusters across cosmic time. Coupling the high spatial resolution of MAVIS ($\sim$0.02$\arcsec$ in the V band) with the high spectral resolution mode (R$\sim$13000) of the proposed spectrograph, science cases for GCs include probing as-yet unresolved cluster cores to examine chemical inhomogeneities and binary and multiple stellar populations \citep[for a complete list of science cases see][]{mavisscience}. A key capability of MAVIS will be to couple chemical abundances and kinematic information with spatially resolved photometry in a homogeneous way. 

A MAVIS science case that will demand both the high spatial and spectral resolution of MAVIS is the search for intermediate mass black holes (IMBHs) in cluster centres. IMBHs, in the mass range of $10^{3}-10^{5}$~M$_{\odot}$ have been proposed as the ``missing link'' between the formation of stellar mass BHs and the super-massive BHs found in galactic centres \citep{rees78, rees84, miller04, ferrarese05, volonteri10}. If IMBHs exist in GC centres, the mechanism behind their creation is still unclear. One theory, ``runaway collisions'', involves the formation and subsequent merging of massive stars in cluster centres \citep{ebisuzaki01, portegies04, gurken04}. Merging occurs between massive binaries and triple systems at the cluster centre at an increasing rate until a supermassive star ($\sim400-190M_{\odot}$) is formed at the centre \citep{sakurai17}. This central star then collapses directly to form an IMBH. Other theories suggest the formation of IMBHs through the repeated dynamical interactions of close binaries composed of stellar-mass BHs and main sequence stars coupled with accretion on longer timescales \citep{giersz15}. 

\begin{figure}
\centering
\includegraphics[scale=0.42]{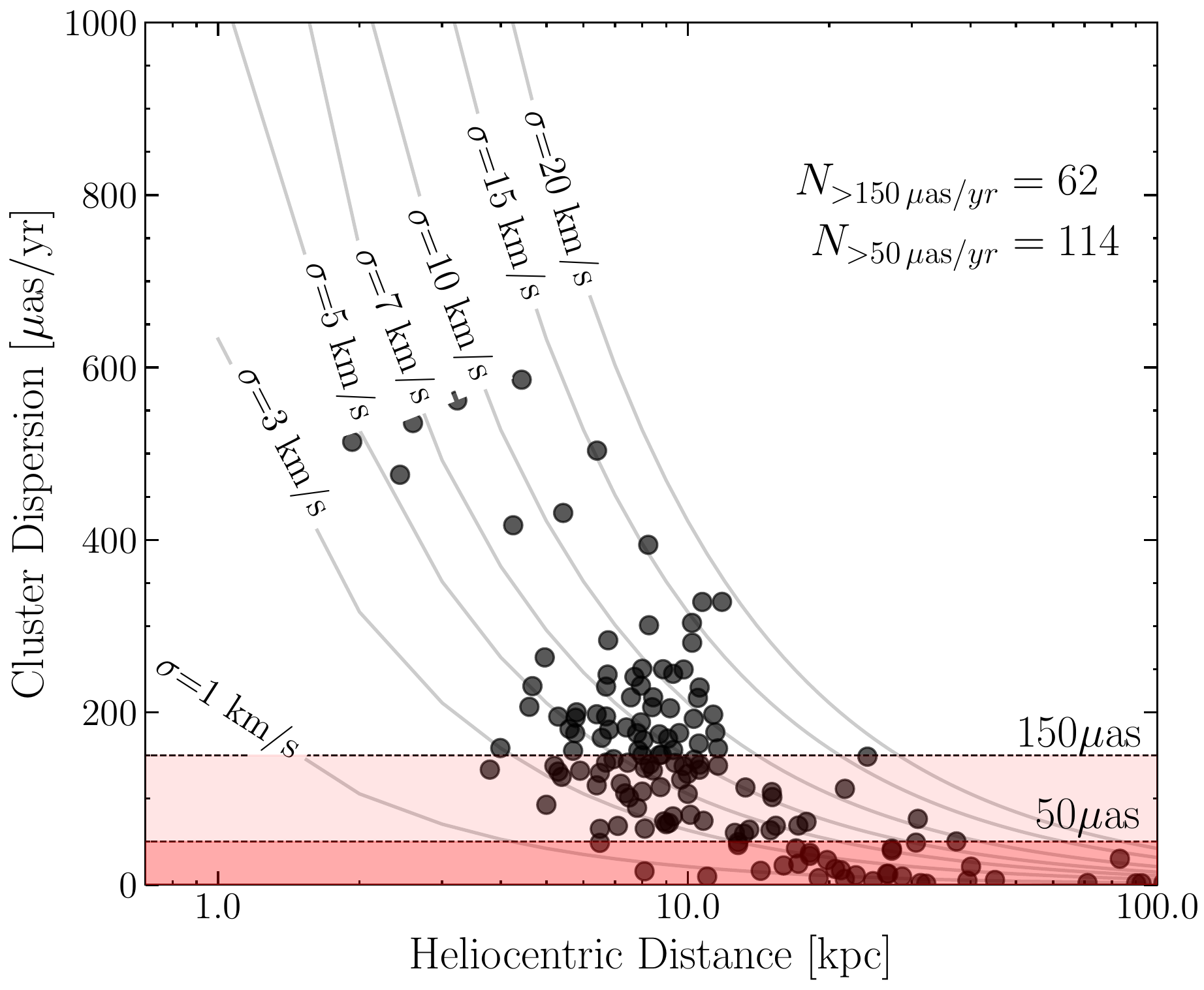}
\caption{\label{fig:mwgcs} Central dispersion of several MW GCs as a function of cluster distance, all values are taken from the catalogue of \citet{baumgardt18}. The shaded regions indicate the requirement and goal for MAVIS astrometric precision (150$\mu$as and 50$\mu$as respectively as marked), giving an indication of IMBH detectability. The number of GCs above these limits are also indicated.
}
\label{fig:gcsample}
\end{figure}

Thus far, observational evidence for a low mass (350~M$_{\odot}$) IMBH in a young, dense star cluster has been found through X-ray observations \citep{portegies04}. But as of yet, no undisputed evidence of IMBHs at the centre of GCs has been found, with many early assertions being challenged through either additional observations and/or re-modeling \citep[e.g. in the case of 47 Tuc;][]{grindlay01, derijcke06, kiziltan17, freire17, brunet20}. From a strictly stellar dynamical point of view, the characteristic signature alluding to the presence of an IMBH is a rise in velocity dispersion in the cluster core. An increase in measured velocity dispersion from PM measurements (subject to considerations of cluster anisotropy), coupled with a cluster surface brightness profile can be used to infer the expected dispersion from visible matter alone \citep[e.g.][]{anderson10, noyola2011}. The discrepancy in observed dispersion can then be associated with a dark matter component, be-it a collection of stellar-mass black holes or a central IMBH \citep{brunet20}. The approach was used by \cite{haberle2021} in their study of the MW GC NGC~6441 using the NACO instrument \citep{naco1, naco2} at the VLT. By coupling their NACO data with existing HST data they recovered PMs over a 15 year epoch with a precision of $\sim30$~$\mu$as/yr for the brightest stars. They concluded that a central IMBH in NGC~6441 could neither be confirmed, nor ruled out finding an upper mass limit of $\mathrm{M}_{\mathrm{IMBH}}<1.32\times10^{4}~\mathrm{M}_{\odot}$. 

Alternatively, direct measurement of the line-of-sight velocities of stars near the cluster core could be used to infer the presence of an IMBH \citep[e.g.][]{kamann16}. To put a detection on firmer footing, both techniques have been used, coupling line-of-sight velocity measurements from one instrument with PM measurements from HST \citep{gebhardt05}. Recently, a survey using the MUSE IFU spectrograph \citep{muse} (without AO), was undertaken of 25 MW GCs in part to examine the internal dynamics \citep{kamann18}. Using a powerful de-blending technique developed in \cite{kamann13}, the survey team were able to extract a total of 200,000 stars. The technique relies on the use of a complementary photometric catalogue of similar spatial resolution to perform source detection and a well described analytical model of the PSF \citep{kamann13}. Using this technique \cite{kamann16} measured the line-of-sight velocities in NGC 6397 for 12,000 stars, concluding that an IMBH \textit{could} be present. Coupling the higher spatial and spectral resolution of MAVIS, a homogeneous study of GC cores could be undertaken with high enough discrimination to say with certainty whether central IMBHs are present. 

Fig.~\ref{fig:mwgcs} shows MW GCs with measured central velocity dispersions as a function of heliocentric distance. The dispersion values were taken from the catalogue of \cite{baumgardt18} and converted to PM. The contours of constant velocity dispersion highlight the increasing difficulty of PM measurements at increasing heliocentric distances. The red shaded regions highlight the MW GCs with central dispersion values that could be resolved by MAVIS given the predicted astrometric precision alone. To detect an IMBH in any of these clusters MAVIS would need to resolve the dynamical signature of the IMBH in addition to the internal dispersion. Note that the number of clusters with resolvable internal kinematics increases by nearly a factor of two if MAVIS achieves its goal of 50~$\mu$as precision. 

Formally, the astrometric requirement for MAVIS is that for high signal-to-noise sources (S/N $>200$) separated by no more than one arcsecond, MAVIS will provide an astrometric accuracy of 150~$\mu$as (with the goal of 50~$\mu$as). Furthermore, this assumes that at least three astrometric (e.g. \textit{Gaia}) reference sources are present in the image and that the lower order plate scale and rotational terms can be calibrated to within 0.01\%. 
Although a highly technical requirement, as a first step to verifying that it can be achieved, several sources of error must be quantified and understood to predict the true capabilities of MAVIS. It is with this goal in mind that we have designed the MAVIS Image Simulator (\texttt{MAVISIM}).

This paper is organised as both a review of the common astrometric error sources present in MCAO systems, presented with observational astronomers in mind, as well as an exploration into the predicted capabilities of MAVIS, the first workhorse visible MCAO instrument. Section~\ref{sec:allterms} presents an initial astrometric error budget for MAVIS and a description of three major astrometric errors considered in this study. Section~\ref{sec:mavissimim} then transitions to a discussion of how we have chosen to model the aforementioned errors in \texttt{MAVISIM}. In Section~\ref{sec:imbh}, drawing inspiration from \cite{fiorentino20}, \texttt{MAVISIM} is used to explore the possibility of detecting a central IMBH in a MW GC with MAVIS. Section~\ref{sec:results} presents the results of an exploration into the predicted dominant source of astrometric error in MAVIS and the results of the IMBH investigation. Finally, Section~\ref{sec:futwork} discusses what we aim to incorporate into the next version of \texttt{MAVISIM}.

\begin{table*}
	\centering
	\caption{The preliminary astrometric error budget for MAVIS including the different sources, their expected effect and impact on astrometry and whether they are included in this first version of \texttt{MAVISIM}. Note that the effects of each error term are characterised as having a high- (HO) or low-order (LO) distortive effect on the PSF. The terms shown in bold are discussed in more detail in section \ref{sec:allterms}.}
	\label{tab:astrobudget}
	\begin{tabular}{lllc}
		\hline
		Error Type & Description	& Expected Impact   & In \texttt{MAVISIM 1.0}\\
		\hline
		\textbf{PSF Field Variability} & HO and LO spatial variability of the PSF across the FoV &  Large  & \checkmark \\
		\textbf{Tip-Tilt Residual Error} & Degradation of the PSF (core) across the FoV & Large  & \checkmark \\
		Atmospheric Differential Refraction & Positional bias after ADC correction & T.B.D. & \xmark \\
		Charge Diffusion in Detector Pixels & Broadening of PSF core & Intermediate & \checkmark \\
		CCD Pixel Quantum Efficiency Inhomogeneity & Intra-pixel and inter-pixel effects & T.B.D, Small & \xmark \\
		Photon \& Read Noise & Overall degredation of PSF & Intermediate & \checkmark \\
		\textbf{Static Distortion from the AO Module} & HO and LO effects to PSF & Small & \checkmark \\
		Dynamic Distortion from the AO Module & HO and LO effects to PSF & T.B.D & \xmark \\
		Telescope Plate Scale Variations & LO effect to PSF & T.B.D. & \xmark \\
		Telescope Field Distortion & HO effect to PSF & T.B.D. & \xmark \\
		AO Module Deformable Mirrors & HO effect to PSF & Small & \xmark \\
		\hline
	\end{tabular}
\end{table*}

\section{The Preliminary MAVIS Astrometric Error Budget}
\label{sec:allterms}
\todo{The preliminary MAVIS astrometric error budget is shown in Table~\ref{tab:astrobudget}. Each error is listed alongside a brief description of its expected impact, scale and whether it is modeled in this first iteration of \texttt{MAVISIM}. The current version of the budget draws from existing astrometric investigations and budgets for other MCAO-fed instruments (\citealp[eg. MAD:][]{madastrom}, \citealp[GeMS/GSOAI:][]{neichel14}, \citealp[TMT/IRIS:][]{tmtastrombud1, tmtastrombud2}, \citealp[ELT/MICADO:][]{trippe2010,micadodist}) with many sources in common. While identifying the absolute scale and impact of each effect is difficult and likely not possible before going on-sky, we aim to incorporate as many of the listed error sources as possible in \texttt{MAVISIM}, presenting a new way to directly test the impact of each on real MAVIS science cases. In the following section we provide a brief description of the three major astrometric errors we have modeled thus far.}

\subsection{\label{sec:psfvar}PSF Field Variability}
Positionally-dependent changes to the PSF profile, characterised as ``PSF field variability'' are both dynamic and static perturbations to the PSF. The primary cause of the variability is anisoplanatism between the science object and AO reference sources. In the case of high-order (HO) spatial variability, stemming from the LGS reference sources, the static contributions are associated with the fixed geometry of the LGS constellation, while dynamic contributions can result from a time-dependent sodium layer profile \citep{sandrine08}. Briefly, static anisoplanatism occurs when the science object and AO reference object (LGS/NGS) do not occupy the same isoplanatic patch and thus occupy regions of non-identical turbulence \citep{fried82}. This results in a worsening AO correction with increasing angular distance from the reference object, manifesting as a spatially-dependent degradation of the PSF.

\subsection{\label{ttres}Tip-Tilt Residual}
Representing the dominant aberrations in all AO systems, uncorrected tip-tilt (TT) terms introduce an unknown shift to the centroid of science objects. Understanding the residual error associated with the expected TT correction is therefore a critical component of our astrometric budget. The TT residual is dynamic in reality due to e.g. the effects of ``windshake'', telescope vibrations and measurement errors associated with NGS flux \citep{kulcsar2012}, but is often modeled as a single, static error. The effect of NGS anisoplanatism also leads to a spatially variable TT residual. In practice, temporal and spatial changes to the TT residual affect differential astrometry by changing the measured distance between science objects across different exposures \citep{cameron09}. Ignoring the contributions from the telescope to the residual, the NGS characteristics and configuration govern the scale and field variability of the TT residual. Selecting an optimum NGS constellation is limited by the availability of guide stars both in terms of geometry and brightness.

\subsection{\label{sec:lowordertheory}Low Order Static Distortion from the AO Module}
Considering the AO module (AOM) only, both static and dynamic distortions contribute to the astrometric budget. Static distortion largely stems from non-common path aberrations (NCPAs) and off-axis alignment of optical elements and can be modeled using ray-tracing software or recovered on-sky using a calibration (pinhole) mask. In the GeMS AOM, the presence of significant static distortion in the system has been a limiting factor for achieving precision astrometry \citep{massari16b}. The static distortion contributes a fixed spatial offset to the stars that can be mis-interpreted as an additional PM term. \todo{To improve astrometric measurements with GeMS post-commissioning, the static distortion has been studied} using both the ray-tracing technique \citep{patti2019} \todo{and an astrometric mask which was installed in 2017 \citep{gemsmask}. Both studies found that $\mu$as distortion residuals were achievable when fitting the distortion pattern with at least a third order polynomial.} 

\section{The MAVIS Image Simulator}
\label{sec:mavissimim}
\texttt{MAVISIM} was designed to produce ``realistic'' MAVIS images that can be used to explore both the technical and scientific capabilities of MAVIS. It is written entirely in \textsc{Python} using an object oriented approach. Following the example set by \texttt{SimCADO} \citep{simcado}, we have bundled \texttt{MAVISIM} as a \textsc{Python} package to simplify its distribution. In the following section we present how each error term included in \texttt{MAVISIM} is modeled. Each of the error terms is modeled independently and can therefore be analysed separately to examine their effect on astrometric performance. In Section~\ref{sec:image_gen} we give an overview of the steps used to generate the final image, the inputs, and the final product.

\subsection{\label{sec:modeling_error_terms}Modeling Astrometric Error Terms}

\subsubsection{\label{sec:psfvarmodel}PSF Variability: High Order Aberrations}
In Section \ref{sec:psfvar} we described PSF field variability due to anisoplanatism in a general sense, in the context of \texttt{MAVISIM} we use the term ``HO PSF field variability'' to describe the effects of anisoplanatism specifically associated with the high-order (HO) LGS corrections. In the modal modeling approach, Zernike polynomials are used to model the wavefront \citep{noll76}. LGS light is used to recover the aberrations represented by the HO Zernike polynomials. Among the HO aberrations, anisoplanatism is the most dominant. In practice, LGS anisoplanatism affects the entire PSF profile but dominates inside the AO control radius ($\sim0.28\arcsec$ at 550~nm). Where the control radius is defined in radians as, $r_{\mathrm{corr}}=\lambda n/2D$, with $n$ being the number of actuators in the telescope pupil diameter (40 for MAVIS), $\lambda$ the image wavelength and $D$ the telescope diameter. 

In this first iteration of \texttt{MAVISIM} we simulate the MAVIS PSF using a Fourier-based method \citep{roddier, jolissaint, neichel2008, neichel2021, agapito2020}. This method has the advantage of modeling the effects of imperfect HO corrections very quickly. Although a more accurate model of the MAVIS PSF has been generated using an end-to-end AO simulation in \textsc{COMPASS} \citep{compass, jessemavis}, we choose to experiment only with the Fourier PSF at this point. The reason for this is two-fold: i) the quick generation of PSFs across the MAVIS FOV on the time scale of seconds and ii) the ability to add the low order (LO) TT residual term separately, providing flexibility on the choice of NGS constellation. Although \texttt{MAVISIM} can accept PSFs generated using either method, we postpone the use of an end-to-end PSF model to a later version of the simulator, discussing the specific test cases in Section~\ref{sec:etepsf}. 

Briefly, the Fourier method provides a representation of the long-exposure, time averaged PSF. The PSF is derived by taking the Fourier transform ($\mathscr{F}$) of the system Optical Transfer Function (OTF) as described by: $\mathrm{PSF}=\mathscr{F}\{\mathrm{OTF_{\mathrm{system}}}\}=\mathscr{F}\{\mathrm{OTF}_{\mathrm{DL}}\times\mathrm{OTF}_{\mathrm{AO}}\}$. Where the system OTF itself is expressed as the product of the OTF of the telescope and instrument in the turbulence-free case ($\mathrm{OTF}_{\mathrm{DL}}$), and the OTF resulting purely from the turbulence residuals after AO compensation ($\mathrm{OTF}_{\mathrm{AO}}$). The AO OTF itself can be computed from the structure function of the compensated phase, the difference between the true and reconstructed wavefront. The latter is estimated from the AO error term including but not limited to anisoplanatism, servo-lag introduced by the AO loop, aliasing associated with the wavefront sensor \citep{rigaut1998, gendron2012} and measurement noise.

\begin{figure}
\includegraphics[width=\linewidth]{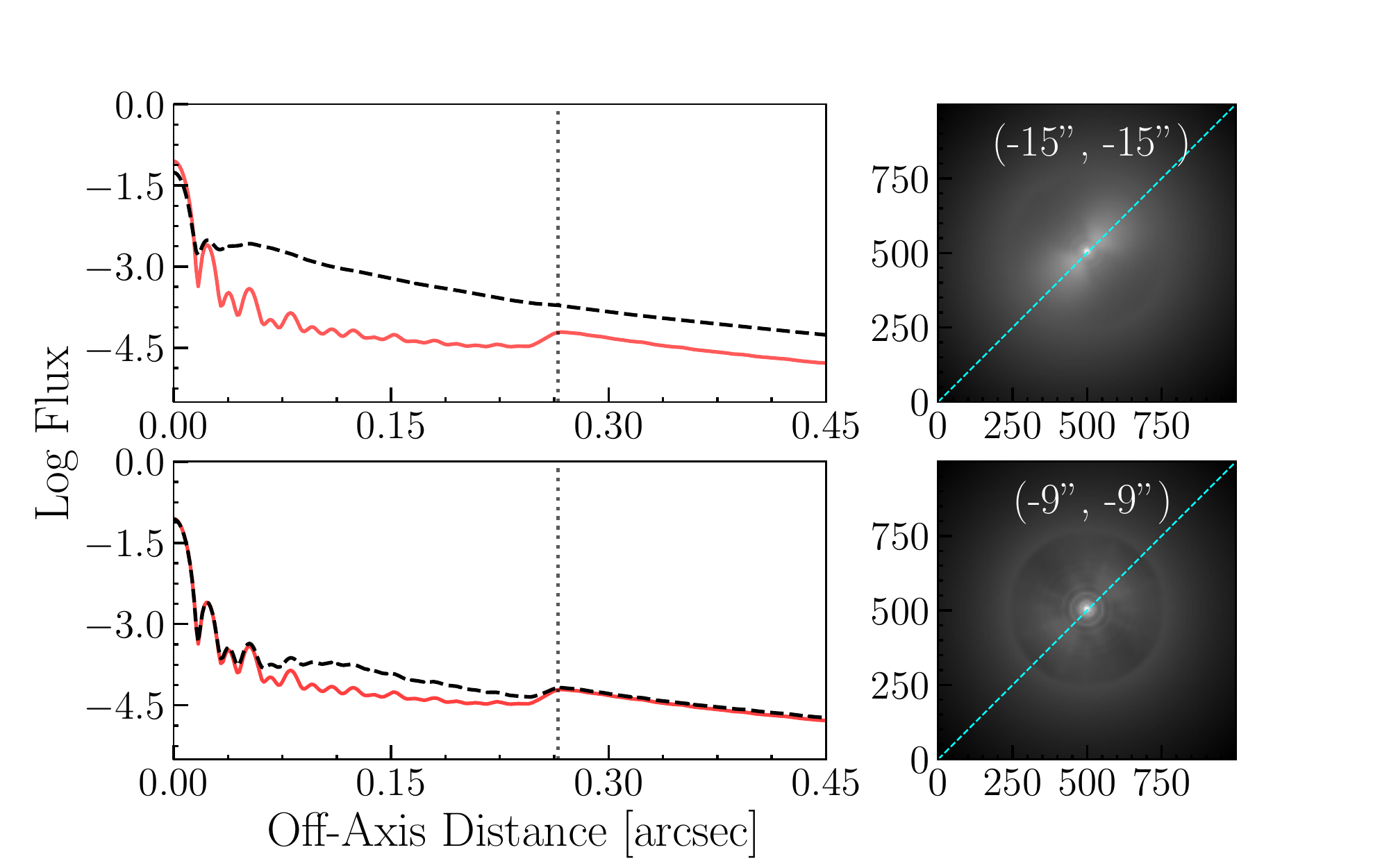}
\caption{Radial profile of the MAVIS PSF at two different points in the imager field, (\textit{top}) at the edge of the field in the lower left corner (-15\arcsec, -15\arcsec), demonstrating the worst case PSF and (\textit{bottom}) slightly inwards from the corner of the field (-9\arcsec, -9\arcsec). Both are shown as black dashed lines. Plotted beneath each profile and shown in red, is the profile of the PSF at the centre of the FoV (0\arcsec, 0\arcsec). The AO control radius is marked in each profile with a black dotted line. To the right of each profile is a log intensity thumbnail PSF, highlighting the radial elongation of the PSF due to anisoplantism. The axes are both in pixels and mas, as the pixel sampling is one pixel/mas.}
\label{fig:psfprofile}
\end{figure}

To cover the range of PSF spatial variability across the 30\arcsec$\times$30\arcsec\ imager field, we create an equally sampled grid of 11$\times$11 PSFs using the Fourier method. A Bilinear interpolation is then used to determine the PSF at interim positions in the grid. To facilitate rapid experimentation of the different error terms, we trim the Fourier PSF to enclose $>95\%$ of the encircled energy. An example profile of the Fourier PSF is shown in Fig.~\ref{fig:psfprofile}, with the AO core (control radius) marked for clarity.  

Two examples of the MAVIS Fourier PSF radial profile are shown in Fig.~\ref{fig:psfprofile} at different locations in the imager FoV. Adopting the convention that the centre of the MAVIS imager is (0\arcsec, 0\arcsec), the PSFs selected are from the lower left hand corner (-15\arcsec, -15\arcsec) and slightly inwards (-9\arcsec, -9\arcsec). Note that the PSF at (-15\arcsec, -15\arcsec) is actually outside of the optimally corrected MAVIS science field, a circle with $r=15\arcsec$. The radial profiles are shown in the left plots alongside log-intensity thumbnails of the PSFs where the radial slice is denoted with a cyan line. Both radial profiles are plotted on top of the radial profile of the central PSF (0\arcsec, 0\arcsec), shown in red. The PSF at (-15\arcsec, -15\arcsec) shows the most obvious disagreement with the central PSF profile, with the bulk of the light pushed out from the central peak and out to radii closer to the AO control radius ($r\sim0.28\arcsec$). In fact, the control radius is not easily discernible in the worst-case PSF in the direction of elongation. The second case, closer to the centre at (-9\arcsec, -9\arcsec) shows better agreement with the central PSF profile, even in the direction of elongation.

\begin{figure*}
\includegraphics[scale=0.55]{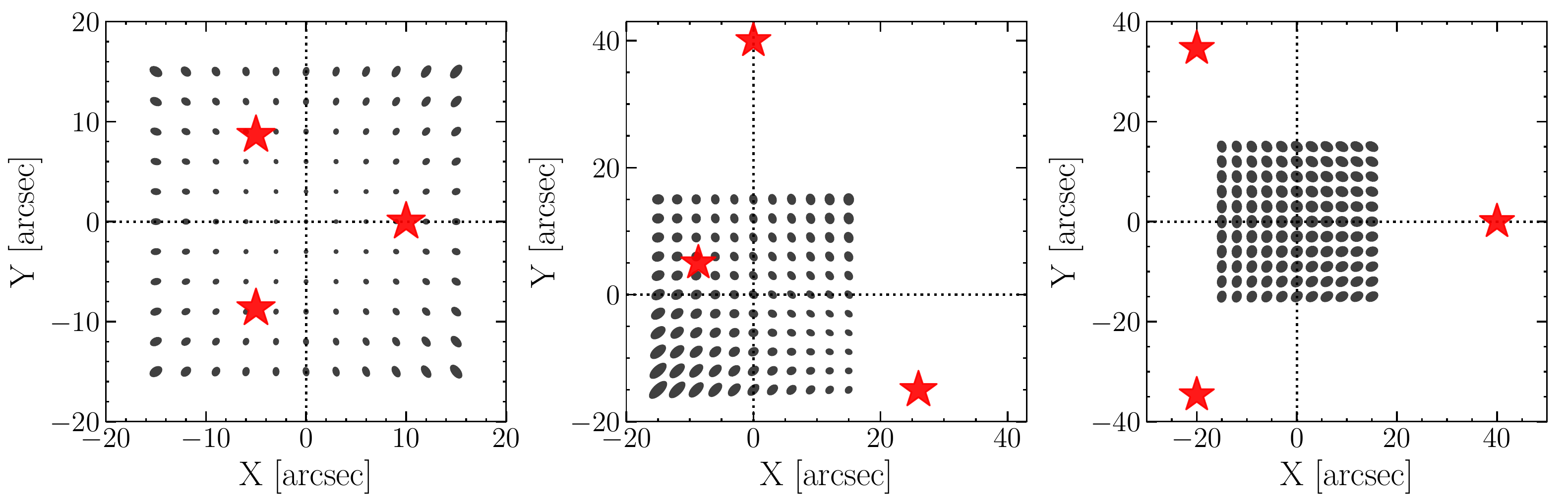}
\caption{Three examples of TT kernel maps created using the method described in Section~\ref{sec:ttresmod} for the case of a ``good'' (\textit{left}), typical  (\textit{middle}) and ``bad'' constellation. All three maps show the residual in arcseconds multiplied by a factor of 100 and the centre of the imager field marked with cross-hairs. Details of the three NGS (shown as red stars) brightness and configurations are given in Section~\ref{sec:ttresmod}.}
\label{fig:ttjitter}
\end{figure*}

\subsubsection{\label{sec:ttresmod}Tip-Tilt Residuals: Low Order Aberrations}
In practice the effect of the TT residual is to broaden the probability distribution function associated with each measured object, smearing and increasing the size of the core of the PSF ($\sim0.18\arcsec$ at 550~nm vs. the diffraction limited case of $\sim0.14\arcsec$). In addition to increased PSF core sizes, anisoplanatism arising from the NGS constellation geometry introduces a spatial dependence on the final shape of the PSF core. Also, as the TT determination is sensitive to measurement noise, which is governed by the number of NGS photons, the magnitude of the TT residual depends on the stellar NGS magnitudes as well.

As the TT residual can be approximated using Gaussian statistics \citep{scot93}, it can be visualised and modeled as a 2D multivariate Gaussian kernel.
The shape and size of the kernel at different points in the field, parametrised as the semi-major axis, semi-minor axis and inclination angle capture the magnitude and field variability of the residual.
The time averaged TT kernel is then calculated at various points in field by adding the total tomographic error in quadrature with the total vibration/windshake error, and a term capturing the noise covariance matrix of the three NGSs as $\sigma_{\mathrm{TT}}^{2}=\sigma_{\mathrm{vib}}^{2}+\sigma_{\mathrm{tomo}}^{2}+\sigma_{\mathrm{noise}}^{2}$. Note that the tomographic error encompasses the anisoplanatic term.

\begin{table}
	\centering
	\caption{Parameters for the three NGS constellations used to derive the TT residual maps. The location of each NGS is listed in polar coordinates with the centre of the imager being (0\arcsec,0\arcsec), the magnitude of the guide star in the $H$-band is listed as the third parameter.}
	\label{tab:ttjitter}
	\begin{tabular}{llll} 
		\hline
		Const. & NGS1    & NGS2  & NGS3\\
		        & ($\arcsec$, $^{\mathrm{o}}$, mag)   & ($\arcsec$, $^{\mathrm{o}}$, mag)  & ($\arcsec$, $^{\mathrm{o}}$, mag)  \\
		\hline
		Good & (10, 0, 15)  & (10, 120, 15) & (10, 240, 15) \\
		Typical & (30, 330, 16)  & (40, 90, 18) & (10, 150, 18) \\
		Bad & (40, 0, 18) & (40, 120, 18) & (40, 240, 18) \\
		\hline
	\end{tabular}
\end{table}

To determine the TT contribution under different conditions, we consider three different TT kernel maps derived using the different NGS constellations given in Table~\ref{tab:ttjitter}. These three maps capture the ``good'', ``typical'' and ``bad'' TT cases. Interestingly, although the magnitude of the TT kernel is the largest in the case of the bad constellation, it is more \textsl{variable} in the case of the good constellation and even more so in the case of the typical constellation. The effects of both the magnitude and variability of the TT kernel are investigated in subsequent sections.

The three maps of the calculated TT kernel, sampled every 3\arcsec across the imager FoV, are shown in Fig.~\ref{fig:ttjitter}. Each point in the maps represents the projected shape of the TT kernel. The NGS constellations are also shown in Fig.~\ref{fig:ttjitter} as the red stars. The left map shows the good constellation configuration, the centre shows the typical configuration and the right the worst configuration. In the the typical map, only NGS3 lays in the science field, while the worst map contains no NGSs in the science field. In the case of Fig.~\ref{fig:ttjitter} and all the simulated constellations, the TT kernel spatial variability is relatively smooth and thus easily interpolated. This allows for the creation of a field-dependent and thus object-dependent TT kernel at each point in the MAVIS imager FoV. Additionally, the Gaussian nature of the kernel allows for sub-pixel positional shifts to be captured in the kernel itself.  

\subsubsection{\label{sec:statdist}Low Order Distortion: Static Term}
The MAVIS AOM has been designed to minimise geometric distortion, primarily through the use of on-axis transmissive optics. As MAVIS operates in the visible, transmissive optics provide optimal throughput and remove the need for off-axis optical elements, the primary contributor to field distortion. Off-axis parabolic mirrors, found in current and future near-IR MCAO systems including GeMS \citep{james03} and NFIRAOS \citep{herriot10} are the primary source of off-axis-introduced distortions. Although in the case of NFIRAOS, two pairs of off-axis parabolic mirrors correct for the introduced distortions \citep{nfiraos2018}. In addition to avoiding the use of off-axis elements, the MAVIS AOM will have a dedicated pinhole mask to calibrate both static and quasi-static (night-to-night) distortions. In a subsequent paper we will explore the characteristics and capabilities of the distortion mask. 

\begin{figure*}
\includegraphics[scale=0.45]{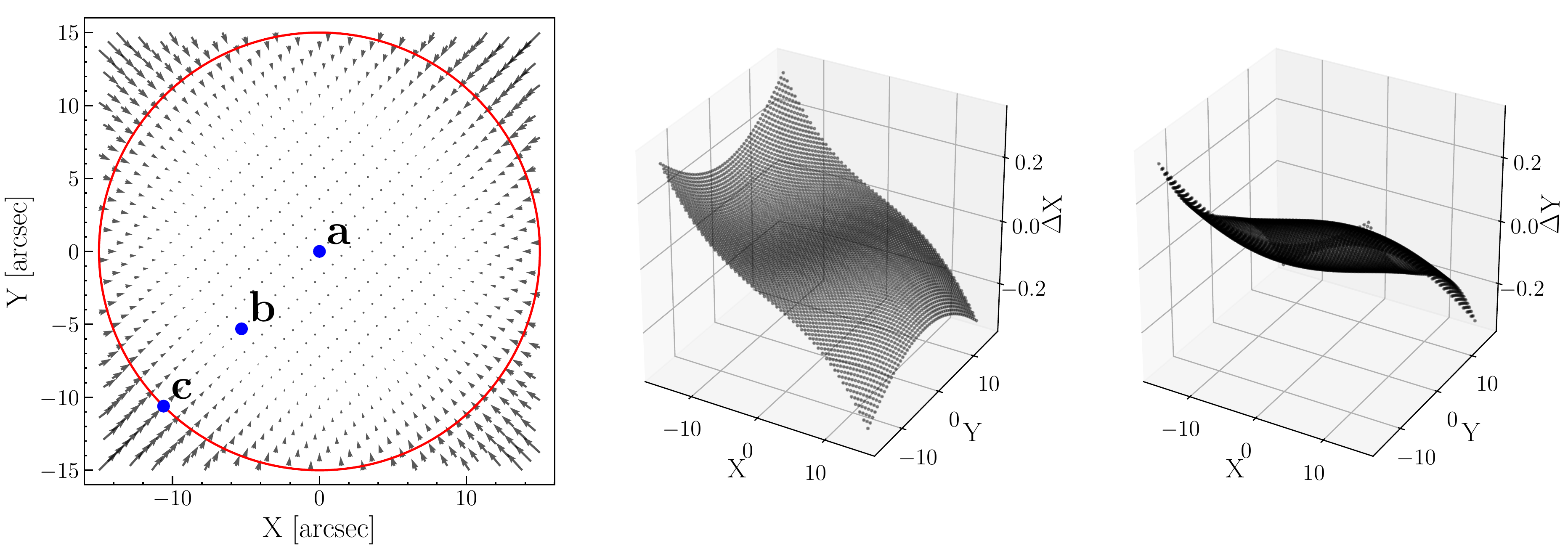}
\caption{Static distortion map and projections created using a ray-tracing method. The MAVIS science FoV is marked as the red circle in the left plot of the static distortion. Three positions are marked in the field as $a$, $b$ and $c$, the distortion values associated with these points are given in Sec.~\ref{sec:statdist}. Projections of the x and y-offsets, scaled by a factor 1000, are included in the middle and right plots respectively.}
\label{fig:staticdistmap}
\end{figure*}

To model the static distortion in the AOM, we use a ray-tracing technique to create a static distortion map (as described in \cite{patti2019}). The map is then interpolated and used to shift each TT kernel. The nominal MAVIS static distortion map is shown on the left in Fig.~\ref{fig:staticdistmap} along with the scaled x and y-projections. The MAVIS science FoV, representing the area with the best AO correction, is shown as the red circle ($r_{\mathrm{SF}}=15\arcsec$) in Fig.~\ref{fig:staticdistmap}. At present we do not consider mid-spatial frequency errors (caused by the grinding/polishing process) for the optical surfaces and therefore expect this to be a lower limit of the true static distortion map. That being said, we see a maximum distortion (summing the x and y contributions in quadrature) of 0.44~mas at (15\arcsec, -15\arcsec) which is outside of the science FoV. For context, \cite{patti2019} recover a maximum static distortion of 180~mas (nine pixels) with a standard deviation of 34~mas (1.7 pixels) across the GeMS field. More importantly, within the MAVIS science field the distortion is significantly smaller. The distortion at points $a$, $b$ and $c$ marked in the distortion map shown in Fig.~\ref{fig:staticdistmap}, is 0~mas, 0.02~mas and 0.16~mas respectively. A further encouraging result is the small value of the standard deviation across the science FoV, 0.05~mas when summing the x and y contributions in quadrature. 

\subsubsection{\label{sec:imagerchar} Modeling the Imager}
The post-focal instrumentation of MAVIS will include a $\sim$4000$\times$4000 pixel imager (the current baseline is a CCD250) to provide diffraction limited imaging of the $r=15\arcsec$ MAVIS science field. Nominally the imager will cover the U to $z$ band, sampling at 7.36 mas per pixel. For simulation purposes we round the pixel size to 7.5 mas per pixel for ease of computation. The true pixel size of 7.36 mas reflects the choice to optimise and over sample around the V band where the FWHM at 550 nm is $\sim17.9$~mas. Slightly larger than the diffraction-limited FWHM, the realistic FWHM is determined by adding the diffraction-limited PSF, $14.2$~mas in quadrature with the average TT residual FWHM from the typical map $8.5$~mas and an additional term due to charge diffusion among adjacent pixels (pixel cross-talk). In the case of the CCD250, we estimate the charge diffusion PSF to be at maximum 94\% of a pixel, or $6.9$~mas. Read out noise and quantum efficiency are based on the detector baseline \citep{rigautmavis}. Additional detector effects like inter-pixel sensitivity variations will be investigated in the future following the method outlined by \cite{chargedif}. The sky background value for Paranal, as well as instrumental and telescope throughput, are included to simulate realistic image characteristics. 

For the case of monochromatic images, we adopt a V band sky brightness of 21.61 mag/arcsec$^{2}$ \citep{patat04}. The throughput of the system is taken into account by combining the expected throughput of the AOM with the average reflectivity of the VLT. This results in a total throughput of $\sim45\%$, which is slightly larger than that of the MAVIS exposure time calculator (ETC) but agrees within 5$\%$ \citep[see][for details of the MAVIS ETC]{mavisscience}. Shot noise is added to the image, assuming the read noise limited case for each pixel (Poisson statistics).  The image is then converted from photons to electrons assuming a quantum efficiency of 89$\%$. Finally, 3~$\mathrm{e^-}$ of read noise is added before capping saturated stars assuming a 16 bit analogue to digital converter.

\subsection{\label{sec:image_gen}Image Generation}
\begin{figure}
\includegraphics[width=\linewidth]{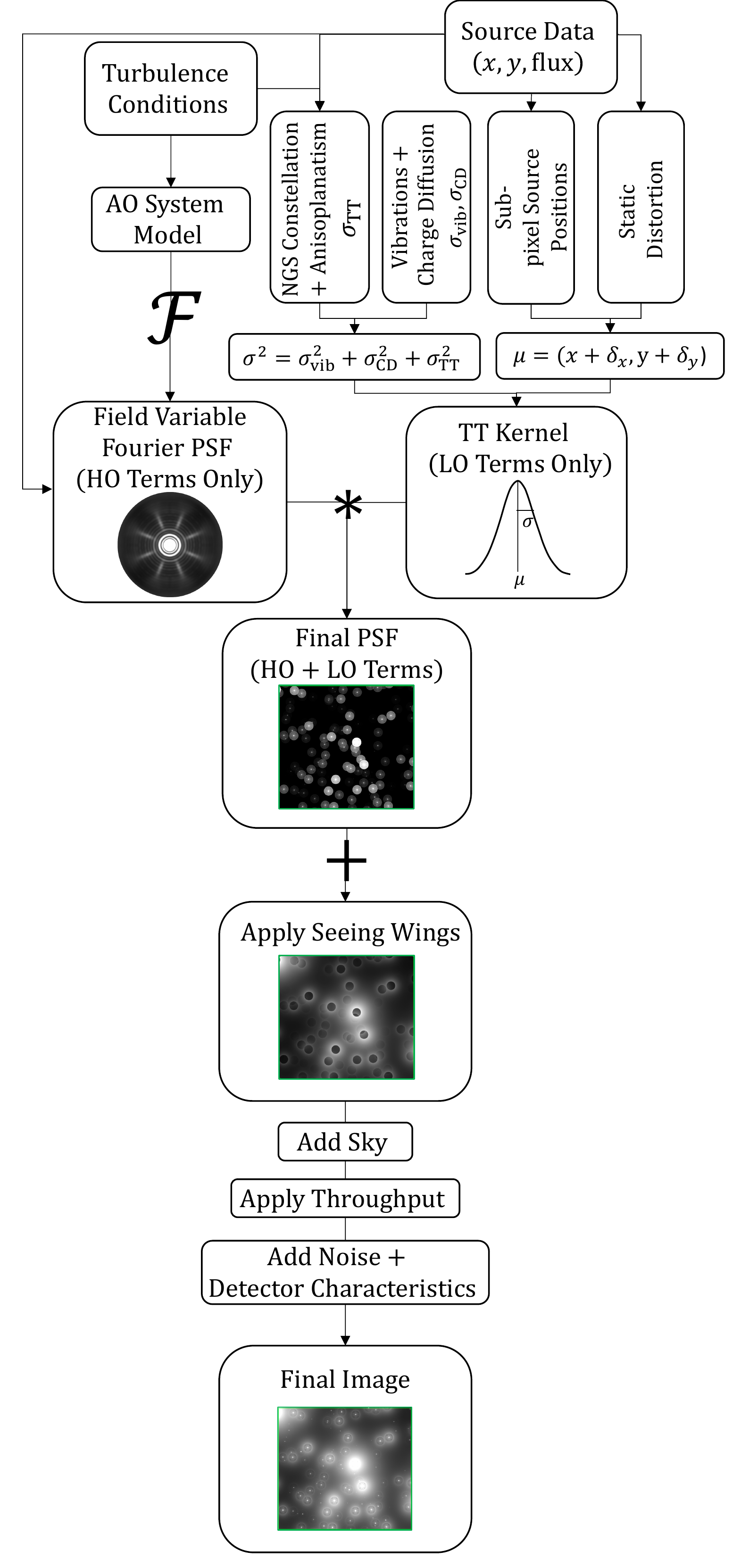}
\caption{Visualisation of the data flow yielding a \texttt{MAVISIM} image. The process is described explicitly in Sec.~\ref{sec:image_gen}. Note that the Fourier transform resulting in each field-variable PSF is denoted as $\mathscr{F}$, the convolution yielding the final PSF for each source is denoted as $*$ and the addition of the seeing wings is marked as $+$.}
\label{fig:mavisim_data_flow}
\end{figure}

Having described each component of \texttt{MAVISIM}, we now provide an overview of the entire process leading to the creation of monochromatic image in Fig.~\ref{fig:mavisim_data_flow}. Beginning with the source data, each object (point source) is represented using it's position and flux. This information is then used in-part to create a unique TT kernel represented by a Gaussian. The mean value of the kernel is derived using the object's positional information including sub-pixel shifts and field-dependent static distortion. The standard deviation of the Gaussian kernel is then calculated by summing the charge diffusion term (3~mas), an additional vibration term (3.5~mas) and the object-dependent TT residual in quadrature. The vibration term is fixed at 3.5~mas to capture a pessimistic estimate of the telescope vibrations and additional unknown terms. The kernel is then scaled using the object's flux. Simultaneously, the objects position is used to create a field-dependent ``thumbnail'' Fourier PSF capturing the specifics of the MAVIS OTF and the HO terms out to the control radius. Recall thumbnail-sized PSFs are selected to avoid lengthy convolutions. The object kernel and Fourier PSF are then convolved to create the final PSF for each object in the data set. 

Following the convolution, the PSF of the object is placed into a $5000\times5000$ pixel array with fixed plate scale to be used in the creation of the final image. For simplicity, we call this large array the ``AO Field". A snapshot of the field can be seen in the ``Final PSF (HO + LO Terms)'' block of Fig.~\ref{fig:mavisim_data_flow} Although not pictured in the figure, the object kernels are also placed into a $5000\times5000$ pixel array we call the ``Kernel Field''. To capture the seeing wings of each object we convolve the Kernel Field with a single large $40\arcsec\times40\arcsec$ PSF taken at the centre of the MAVIS FoV. Prior to convolution, information within the control radius of the large PSF is removed and ramped to smooth the transition into the wings. Convolution with the Kernel Field results in the ``Seeing Field'', an image of seeing wings, as shown in the ``Apply Seeing Wings'' block of Fig.~\ref{fig:mavisim_data_flow}. To create the penultimate, noise-free image, we add the AO Field and Seeing Field together. Note that flux is preserved as there is no overlap between the images aside from the small ramped region associated with the central PSF and wings. The sky, noise and detector characteristics are then applied following the procedure outlined in Sec.~\ref{sec:imagerchar} to create the final \texttt{MAVISIM} image. 

\section{Application of \texttt{MAVISIM}: Science Test Case}
\label{sec:imbh}
To test the predicted astrometric capabilities of MAVIS on a real science case, we have simulated multi-epoch observations of the MW GC NGC~3201 to recover the dynamical signature of a central IMBH through measured PMs. In this section we discuss the input N-body catalogue to \texttt{MAVISIM} and the method by which the PMs are extracted. 

\subsection{\label{sec:nbodycat}Simulating a MW-like Globular Cluster}
As input to \texttt{MAVISIM} we adopt the cluster-specific N-body catalogues of \cite{baumgardt17} and \cite{baumgardt18}. As described in \cite{baumgardt18}, the cluster-specific N-body model is chosen from a grid of 2700 simulations created through varying the initial density profile, half-mass radius, cluster metallicity and mass fraction of the IMBH. The best-fit model is then selected by comparing the observed cluster velocity dispersion, surface brightness profile and mass function against the simulated catalogues. We model a ten year epoch and create two sets of input N-body catalogues, each normalised to contain the same distribution of stellar magnitudes and the same total number of stars. The number of stars chosen represents the typical stellar density of the core of NGC~3201 to match the predicted MAVIS imager FoV. One catalogue set contains a central 1500M$_{\odot}$ IMBH, representing $\sim1\%$ of the cluster mass, while the other has no central IMBH.

\begin{figure*}
\includegraphics[scale=0.75]{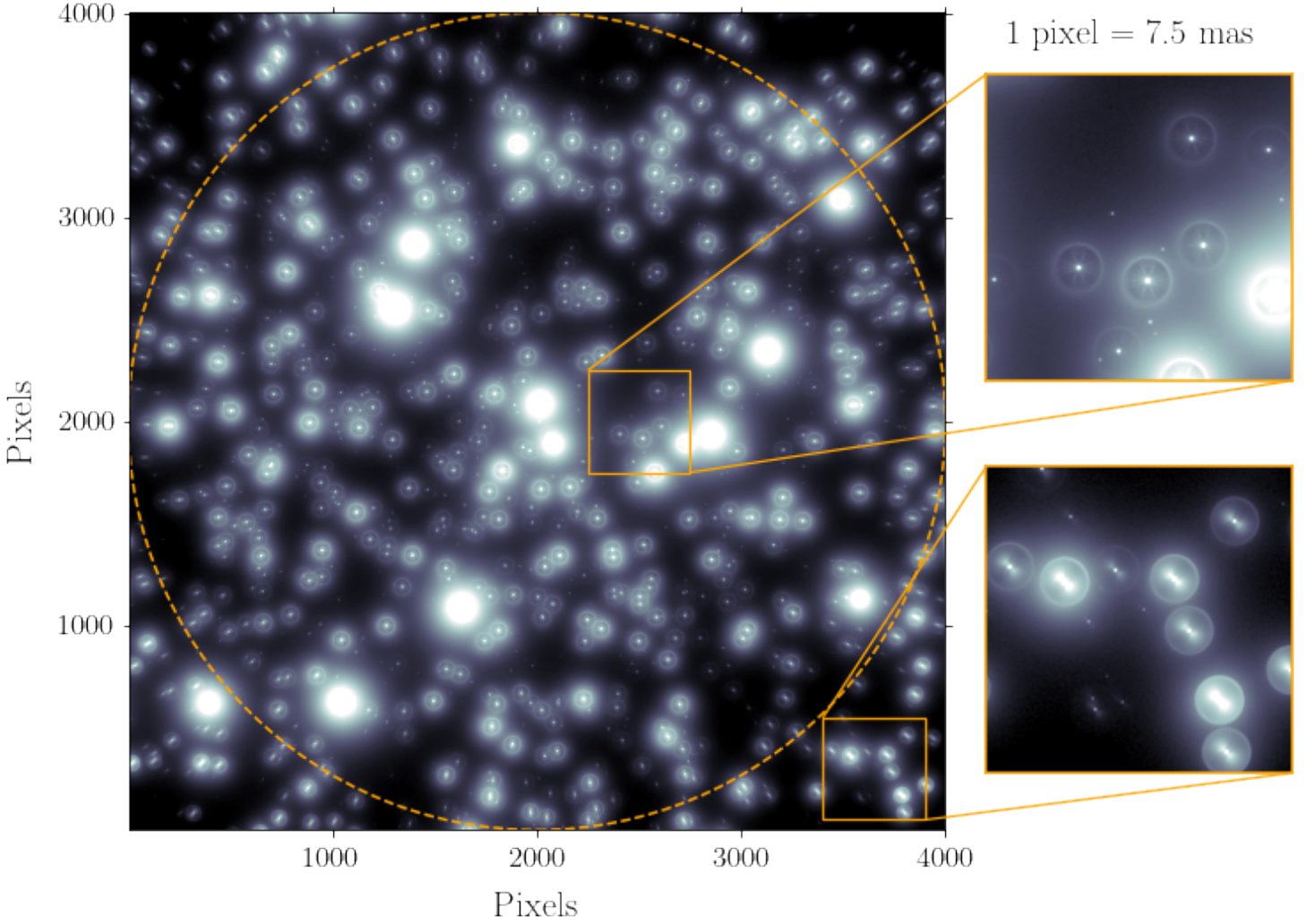}
\caption{Log intensity image created using \texttt{MAVISIM} of the MW GC NGC~3201 integrated for 30s with a pixel size of 7.5~mas (imager $\mathrm{FoV}=30\arcsec\times30\arcsec$). The science FoV ($r=15\arcsec$) is shown as the dotted orange circle to highlight the optimally corrected field. Zoomed insets are shown to demonstrate that \texttt{MAVISIM} is capturing the PSF spatial variability across the imager FoV.}
\label{fig:fullimage}
\end{figure*}

In the simulation we include all the effects discussed in Section~\ref{sec:mavissimim}, although the static distortion is held constant across epochs and thus has no effect on the \todo{simulated} astrometry. \todo{This is a strong assumption and is likely not physical. However, we expect that with the astrometric calibration mask discussed in Section~\ref{sec:calibmask} we will be able to reduce the static distortion present in MAVIS to a negligible level at any epoch.} The turbulence conditions are also kept constant across the two epochs. We select the best case TT residual represented by the ``good'' NGS constellation. This choice is justified by NGC~3201 being a bright and nearby cluster that is likely to contain three $m_{\mathrm{H}}\sim15$ stars in its centre. The flux associated with each star is calculated by \texttt{MAVISIM} using the V band magnitude from the input catalogue. We assume the observations are made at zenith to adopt an extinction of 0.129 magnitudes and calibrate to the Vega photometric system. The total integration time of the simulated images is 3600s, created by stacking 120, 30s frames. The final stacked image is normalised such that the maximum value is the saturation value (65535 ADU). An example image is shown in Fig.~\ref{fig:fullimage}. The HO PSF field variability at two regions in the imager field is highlighted using the zoomed insets, one inside and one outside the science FoV (area of optimal AO correction).

\subsection{\label{sec:astrometry} Recovering Astrometry with \texttt{DAOPhot}}
To recover astrometry in \texttt{MAVISIM} images we use \texttt{DAOPhot-IV} and \texttt{ALLSTAR} \citep{daophot1, daophot2}, standard photometry tools used for MCAO images \citep{turri15, massari2015, massari16a, monty2018, saracino15, saracino16, saracino19}. We follow the standard ``recipe'' for photometry with \texttt{DAOPhot-IV}; i) \texttt{PHOT} to perform initial forced aperture photometry using an input coordinate file ii) \texttt{PSF} to build a mixed analytical-empirical model of the PSF and iii) \texttt{ALLSTAR} to perform the final PSF photometry and re-centroiding. In order to \textit{only} probe the capabilities of PSF-fitting photometry in the extraction of astrometry, we perform forced photometry initially. Forced photometry is standard practice in crowded field and is typically done using an instrument of comparable spacial resolution \citep[e.g. using HST][]{monty2018}. As MAVIS will deliver higher spatial resolution than HST, forced photometry may not be possible in practice, however we accept this assumption in this study to avoid biasing ourselves by the abilities of source detecting software. We use a cubically variable PSF, modeled as Moffat function of $\beta=3.5$, selected following some experimentation. As we choose not to investigate the predicted photometric precision of MAVIS in this paper, we do not optimise \texttt{DAOPhot-IV} for \texttt{MAVISIM} images at this time. We consider this an appropriate compromise as astrometry is most sensitive to modeling of the PSF core and less-so to potentially complex PSF wings. To this end, we evaluate the entire MAVIS FoV as one, selecting stars across the field to build the PSF, rather than performing photometry ``chip-by-chip'' as discussed in \citep{saracino15, turri15}. 

A challenging aspect of the MAVIS Fourier PSF is the complex structure within the AO control radius. Within the control radius, the Fourier PSF displays a pinwheel shape related to the LGS geometry. As such, we experimented with the ideal parameters from which to generate the model of the PSF within \texttt{DAOPhot-IV}. The resulting astrometric accuracy and residual image were used to converge on the best modelling parameters. Ultimately, we adopted an analytical profile within the radius of the first Airy ring, specified using the \texttt{FITRAD} parameter. Beyond the first ring and up until the final discernible Airy Ring we adopted an empirical model of the PSF captured in a lookup table and specified using the \texttt{PSFRAD} parameter. Although we did not examine the photometric results using these parameters, the residuals captured in the subtracted image produced by \texttt{DAOPhot} were satisfactory. In the future, we hope this approach will aid in the analysis of diffraction-limited images with \texttt{DAOPhot-IV} which has proven a robust and capable tool for non-AO and AO images alike. 

Ultimately, an accurate model of the PSF provided alongside observed data will greatly ease the challenges associated with PSF-fitting. A relatively recent and encouraging technique is PSF-reconstruction \citep{veran1997}, where information from the wavefront sensor is used to reconstruct the PSF. It has successfully been used to model the PSF on extreme AO and ground-layer AO systems \citep{beltramo2020, fusco2020} and has already been used successfully as a PSF input to extract highly precise stellar positions and magnitudes \citep{massari2020}.

\section{Results}
\label{sec:results}

In the following section we present the results of simulating the performance of MAVIS including; astrometric precision under different observing conditions and the predicted performance in achieving a key science case, the detection of IMBHs in GC centres.

\subsection{\label{sec:astrobudget}Astrometric Accuracy: Towards a Complete Astrometric Error Budget}
As a reminder, \texttt{MAVISIM} currently models the NGS TT residuals, including the spatially-dependent terms, a field variable PSF capturing LGS anisoplanatism and additional HO terms, and the fixed static distortion from the AOM. Using \texttt{MAVISIM} we can tune each term individually to propagate the effects through to proposed science cases. In this section we examine the consequences of adopting the best and worst NGS constellations and a HO and LO spatially variable PSF on astrometric accuracy. As we assume that the static distortion will be modeled out using our calibration mask and we have not yet modeled a realistic dynamical term, we neglect these terms in our discussion. 

\begin{table}
	\centering
	\caption{\label{tab:simparams} Parameters used, and effects included for each of the eight, 30s simulations. $\sigma_{\mathrm{TT}_{\mathrm{ave}}}$ refers to the TT kernel atmospheric residuals only, while the average FWHM additionally includes the charge diffusion and vibration terms. The fraction of \texttt{DAOPhot-IV}-recovered stars relative to input stars is given in the last two columns for stars brighter than 22 magnitudes and fainter than 22 magnitudes respectively.}
	\label{tab:calibchar}
	\begin{tabular}{lllllll} 
		\hline
		Sim. & NGS  & HO PSF   & $\sigma_{\mathrm{TT}_{\mathrm{ave}}}$ & FWHM$_{\mathrm{ave}}$	& $<\mathrm{m}_{22}$   & $\geq \mathrm{m}_{22}$ \\
	 &				& 		& [mas]									 & [mas]		& 
		\\\hline	
		I 	 & None    & Stat.		& 0 	& 17.99	& 70\% & 28\% \\
		Ia   & None    & Var.       & 0     & 17.99 & 70\% & 28\%\\
		Ib   & Typ.    & Stat.    & 6.45  & 22.20 & 69\% & 28\% \\
		II   & Good    & Var.	& 2.92 	& 18.96	& 69\% & 27\% \\
		III  & Typ.    & Var.  & 6.45  & 22.20 & 69\% & 25\% \\
		IV   & Bad	   & Var.   & 9.23  & 26.58 & 69\% & 22\% \\
		III$\times3$  & Typ.	& Var.	& 19.35	& 42.99 & 68\% & 18\% \\
		IV$\times3$   & Bad		& Var.  & 27.69 & 61.42 & 68\% & 12\%
		\\\hline
	\end{tabular}
\end{table}

To examine the aforementioned effects we generated eight images (simulations) of 30s each. The parameters for each simulation are given in Table~\ref{tab:simparams}. For each simulation listed in Table~\ref{tab:simparams}, we represent the TT residual term derived from each map as the average semi-major axis of the TT kernel, $\sigma_{\mathrm{TT}_{\mathrm{ave}}}$. In every case the final TT kernel is derived as described in Sec.~\ref{sec:image_gen}.

As listed in Table~\ref{tab:simparams}, simulations II-IV have different input TT kernel maps, accounting for the good, typical and bad NGS constellations described in Sec.~\ref{sec:ttresmod} and shown in Fig.~\ref{fig:ttjitter}. In the case of Simulation I, we do not include the NGS TT residuals and instead adopt a static TT kernel comprised only of charge diffusion and vibration terms (see Sec.~\ref{sec:image_gen}). Simulation Ia explores the effects of turning on the HO PSF variability, while holding the TT residual kernel constant. Simulation Ib explores the opposite, keeping the HO PSF static and turning on the TT residual variability by adopting the residual map associated with the typical NGS constellation. Simulations III$\times3$ and IV$\times3$ utilise the same NGS constellation as simulations III and IV but the TT residual term is multiplied by a factor of three to probe the extreme cases. In all cases, except Simulation I and Ib, we adopt a field variable HO PSF. In the case of simulations I and Ib, we adopt the PSF at the centre of the FoV as our static HO PSF, thus neglecting all anisoplanatic terms. At this time we consider only a monochromatic PSF at 550~nm. All images include sky background and detector effects. As input to \texttt{MAVISIM} we use the catalogue described in Sec.~\ref{sec:nbodycat} emulating the MW GC NGC~3201 containing $\sim3500$ stars.

To recover the stellar positions for each simulation we use \texttt{ALLSTAR} as described in Section \ref{sec:astrometry}. Briefly, to model the PSF we select $\sim100$ stars across the FoV and keep the \texttt{DAOPhot} parameters constant across simulations Ib and II-IV except in the case of simulations III$\times3$ and IV$\times3$ where we double the \texttt{PSFRAD} parameter. For simulations I and Ia we model the PSF as both a static PSF and cubically variable PSF and adopt the better results of the two (quantified as smaller astrometric errors, though the results are comparable in both cases). The stellar positions are re-determined during our final run of \texttt{ALLSTAR} by selecting the re-centroiding option and then matched with the input catalogue using the \textsc{SciPy} implementation of a cKDTree \citep{scipy}. When performing \texttt{ALLSTAR} we leave the FITRAD parameter free for each simulation, experimenting with different values to extract the best astrometry. Prior to discussing the results, we remind the reader that the magnitudes discussed throughout this section are the ground truth magnitudes taken directly from the N-body input catalogue. We do not quote the \texttt{DAOPhot} recovered magnitudes at any point. 

\begin{table*}
	\centering
	\caption{\label{tab:simresults} Results of the eight simulations described in Sec.\ref{sec:astrobudget}, showing the average astrometric dispersion at different magnitude intervals. The number of stars used to calculate each average is shown next to the value in parenthesis, if the number of stars is the same as Sim.I (row one) it is not shown. Recall the exposure time for each simulation is 30s. The dispersion is calculated following three iterations of $2\sigma$-clipping Gaussian filtering ($\sim5$\% of stars removed on average). Finer dispersion intervals are given visually in Figs.\ref{fig:astromresiab}-\ref{fig:astromresx3}. Note that in the case of Sim. Ia, Ib and II the brightest bins show large dispersion values due to the measurement of saturated stars.}
	\begin{tabular}{llllllllll} 
		\hline
		Sim. 	& m<16   & m$\in[16, 17)$ & m$\in[17, 18)$ & m$\in[18, 19)$	& m$\in[19, 20)$	& m$\in[20, 22)$	& m$\in[22, 24)$	& m$\in[24, 26)$	& m$\geq26$  \\
		    & [$\mu$as] & [$\mu$as] & [$\mu$as] & [$\mu$as]	& [$\mu$as] & [$\mu$as] & [$\mu$as] & [$\mu$as]	& [$\mu$as]
		\\\hline	
		I 	 & 46.52 (18)  & 71.31 (19)  & 53.19 (81) & 48.62 (144)	& 43.94 (143) & 109.16 (297)	& 291.41 (379)	& 771.35 (205)	& 3347.11 (128) \\
		Ia   & 77.34   & 66.58  & 48.57  & 53.61 & 63.10 & 95.20	& 307.12 (377)	& 977.75 (200)	& 3234.32 (146) \\
		Ib   & 142.64  & 69.35  & 56.38	 & 55.40 & 91.90 (141)	& 127.51	& 423.99 (380)	& 1407.16 (197)	& 2702.02 (131) \\
		II   & 72.99   & 33.63  & 35.49  & 55.81 & 57.41	& 103.18	& 320.64 (378)	& 1431.62 (191)	& 2879.16 (116) \\
		III  & 42.22   & 34.03  & 42.56  & 52.63 & 73.77  & 152.97	& 438.74 (376)	& 1618.90 (179)	& 3447.96 (83) \\
		IV   & 46.96   & 39.44  & 35.55  & 46.62 & 79.91  & 181.75 (296)	& 576.50 (372)	& 2155.87 (155)	& 2894.38 (36) \\
		III$\times3$   & 27.75 (17) & 133.10  & 78.34  & 108.12	& 164.58 	& 364.36 (293)	& 1237.78 (345)	& 3253.26 (98)	& 3976.99 (21) \\
		IV$\times3$   & 68.72 (17) & 61.05 & 138.12 & 163.37 & 270.93 (142)	& 708.75 (288)	& 2388.81 (264)	& 2953.99 (29)	& 2732.31 (19)
		\\\hline
	\end{tabular}
\end{table*}

\begin{figure}
\includegraphics[width=\linewidth]{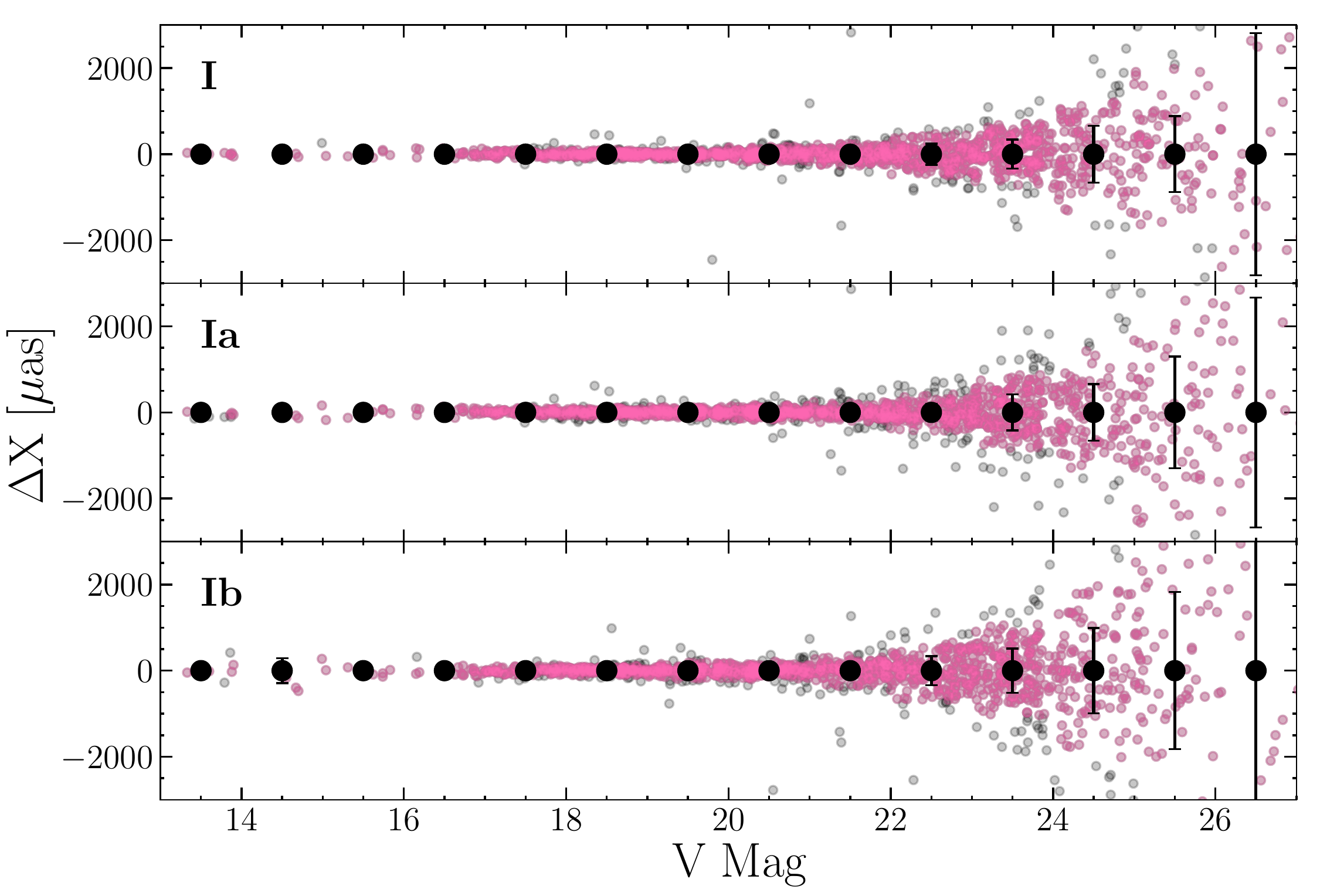}
\caption{Astrometric accuracy as a function of magnitude for simulations I, Ia and Ib. The average dispersion for each one  magnitude wide bin is shown as the black error bars. Note that the average dispersion for each bin was calculated following three iterations of $2\sigma$-clipping Gaussian filtering. Stars removed from the averaging are shown in black, while stars used in the determination of the average are shown in pink.}
\label{fig:astromresiab}
\end{figure}

\begin{figure}
\includegraphics[width=\linewidth]{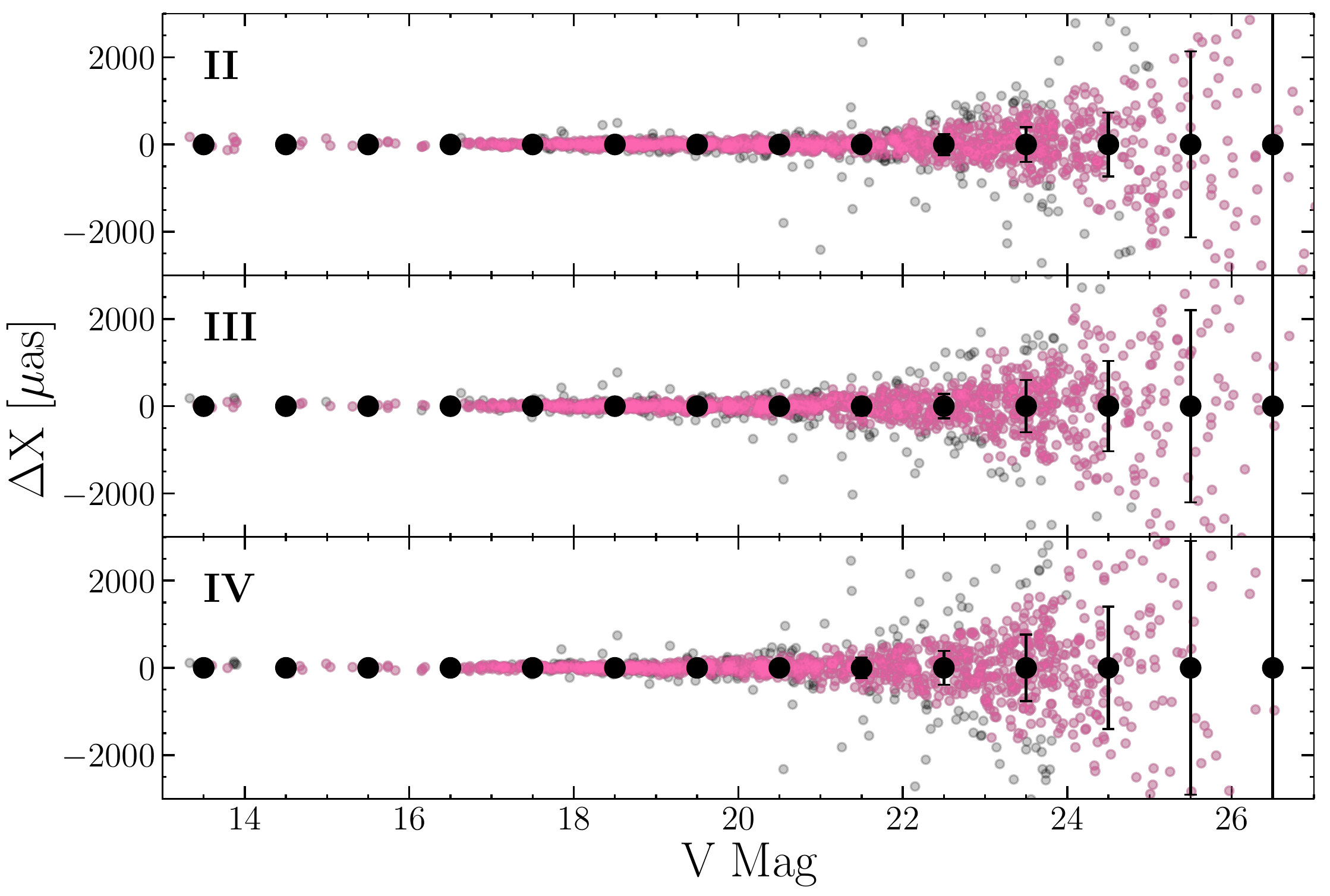}
\caption{Astrometric accuracy and average dispersion as a function of magnitude for simulations II, III and III.}
\label{fig:astromresiiiv}
\end{figure}

\begin{figure}
\includegraphics[width=\linewidth]{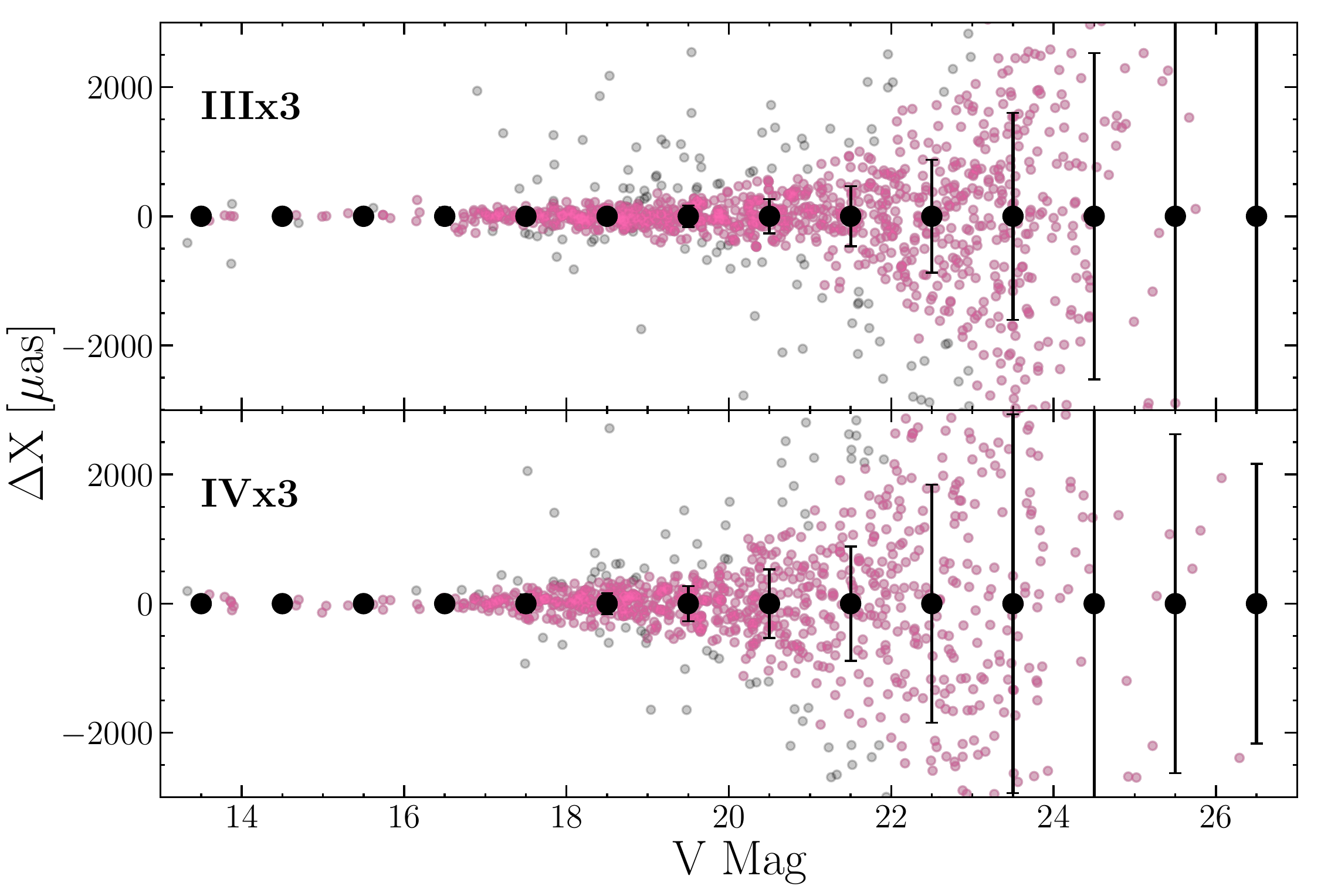}
\caption{Astrometric precision and average dispersion as a function of magnitude for simulations IIIx3 and IVx3.}
\label{fig:astromresx3}
\end{figure}

The results of the eight simulations are shown in Figs.~\ref{fig:astromresiab}-\ref{fig:astromresx3} as the difference in the input $x$-position vs. the recovered $x$-position of each star in the catalogue. The average astrometric dispersion for one-magnitude-sized bins are also given, following three iterations of $2\sigma$-clipping Gaussian filtration. Average dispersion values are also quoted in Table~\ref{tab:simresults}, in slightly wider bin intervals. The number of stars in each bin is included next to the dispersion values. Beginning with simulations I, Ia and Ib we examine the effects of HO PSF and TT residual spatial variability independently using Simulation I as our reference point. Examining both Fig.~\ref{fig:astromresiab} and rows 1-3 in Table~\ref{tab:simresults}, we can see that all three simulations show similar values of dispersion until around $m\approx22$. Above $m\approx22$ the dispersion increases between the three simulations, with the largest increase in dispersion associated with Simulation Ib. Recall Ib includes a spatially variable TT residual kernel and a \textit{static} HO PSF. From this, we can conclude that at faint magnitudes the effects of a field variable HO PSF, dominated by LGS anisoplanatism, are less important than a spatially variable TT kernel. If we consider that the majority of the change to the PSF shape attributed to the HO terms (seen for example in Fig.~\ref{fig:psfprofile}) occurs beyond the first Airy ring and assert that the centroiding algorithm performed through PSF-fitting photometry is most sensitive to the PSF core, this result is not unexpected. Note that this result, as well as those that follow were found to be insensitive to exposure time, as we experimented with increasing the exposure time by a factor of 10 and found the trends remained.

Perhaps the most interesting result is related to the change of NGS guide star constellations, between simulations II-IV as shown in Fig.~\ref{fig:astromresiiiv}. Recall that the three NGS constellations are shown in Fig.~\ref{fig:ttjitter}. Simulations III and IV allow us to probe the effects of the spatial variability vs. the total magnitude of the TT residual. The bad NGS constellation used in Simulation IV displays a relatively smooth residual across the field, as the stars are distributed equidistant from each other. However, the stars are faint and distant from the field centre, leading to the larger overall magnitude of the TT residual. In the case of Simulation III and the typical constellation, both the scale and the orientation of the TT kernel change drastically across the FoV, as can be seen in both plots in Fig.~\ref{fig:ttjitter}. In the bottom left corner of the typical constellation for example, the kernel is elongated in the radial direction relative to the centre of the field, in the top right it is elongated tangentially. 

Examining the results of simulations II-IV shown in Fig.~\ref{fig:astromresiiiv} and summarised in Table~\ref{tab:simresults} we see that the average dispersion increases between simulations II-IV. Again, this disagreement is larger at fainter magnitudes, with the average dispersion nearly doubling between simulations II and IV below $m\approx22$. Fig.~\ref{fig:astromresx3} looks at the difference in dispersion values between simulations III and IV further by amplifying the magnitude of the TT residual kernels. In both Fig.~\ref{fig:astromresx3} and in Table~\ref{tab:simresults} the dispersion is larger in all bins in the case of Simulation IV. Note that in the case of these simulations and some of the others, the brightest bins show large dispersion values due to the measurement of saturated stars. At the faintest end of Simulation IVx3 ($m\geq26$), the completion drops off substantially, explaining the smaller average dispersion. Thus, from simulations III and IV we can conclude that the scale of TT residual field variability can be captured relatively well with a cubically variable PSF model, given that Simulation III is more accurate in general than Simulation IV. In other words, the magnitude of the residual appears to matter more than the spatial variability. The other obvious result between all the Simulations, is the loss of stars at fainter magnitudes in the case of larger TT residual. This is to be expected if we consider that a larger TT kernel has the effect of smearing the light out in the PSF, making the core less sharp. This smearing makes the detection of fainter stars more difficult.

\begin{figure}
\includegraphics[width=\linewidth]{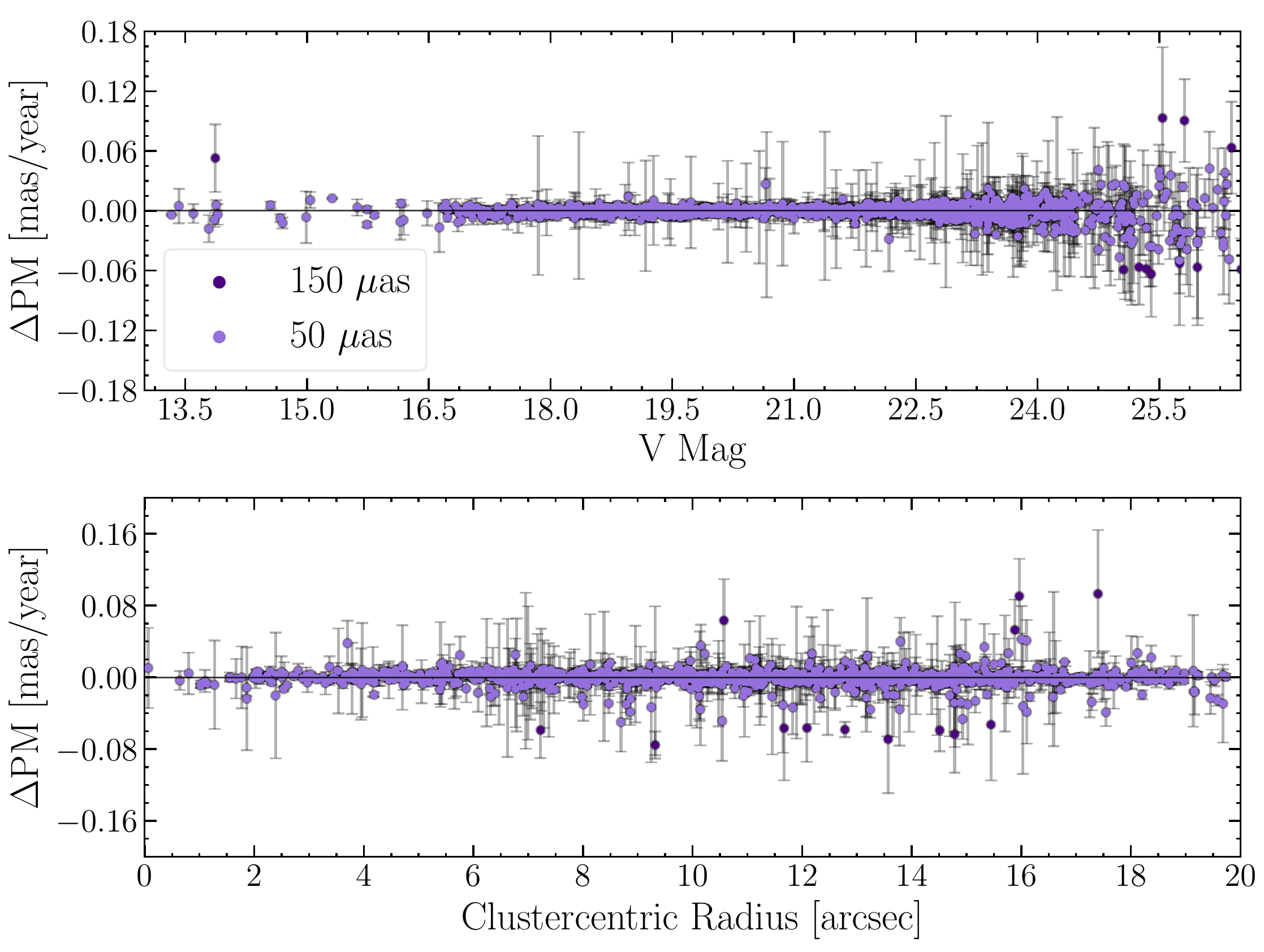}
\caption{Proper motion accuracy represented as the difference between the input and measured proper motions in the $x$-direction as a function of input V-band magnitude. The catalogue has been filtered to remove 147 poor matches ($\sim8\%$ of the original catalogue). The two top-level astrometric requirements for MAVIS are shown as the dark indigo ($150~\mu$as) and purple ($50~\mu$as) points respectively. Simulated results support MAVIS achieving $50~\mu$as accuracy down to relatively faint magnitudes.}
\label{fig:pmresults}
\end{figure}

To compare these results to analogous observational studies, we turn to works examining MW GCs using the GeMS MCAO system. However, when examining the results presented in these studies, it is difficult to decouple the effects of TT spatial variability and the magnitude of the TT kernel. This is because the bulk of the studies examined \citep[e.g.][]{massari16a, massari16b, dalessandro16, monty2018} involve observations with NGS constellations most like our good constellation (the left map in Fig.~\ref{fig:ttjitter}). Therefore, we will try to compare the results indirectly. 

\cite{massari16a} determines the absolute age of the MW GC, NGC 2808 using \texttt{DAOPhot} to perform PSF-fitting photometry. In Fig.~1 of their study they show the FWHM variation across the GSAOI field alongside the location and brightness of the three GeMS NGSs, similar to our Fig.~\ref{fig:ttjitter}. The worst FWHM they measure is associated with chip 4, the most distant chip relative to the NGS constellation and the closest chip to the faintest NGS. They find this directly correlates with the photometric completeness, with chip 4 dropping below 50\% completeness around magnitude 18 in $K_{s}$. If we consider the FWHM to be a proxy for the magnitude of the TT residual, this is generally in agreement with our findings but lacks consideration of the TT variability.

In \cite{monty2018}, a similar study was performed to determine the absolute ages of two MW GCs. They also considered the spatially variability of the GeMS PSF by examining FWHM and ellipticity changes across their images. Again, we can consider the FWHM as a proxy for TT residual magnitude and ellipticity as a proxy for variation in the TT residual kernel across the FoV. They found the cluster with the larger dispersion (NGC~3201) in FWHM ($\sim0.12\arcsec$) across the image showed a smaller dispersion in ellipticity ($\sim0.2$) when compared to the other cluster (NGC~2298). This equates to large overall magnitude and little spatial variation of the TT residual, most like our bad NGS constellation map. The other cluster (NGC~2298) showed a smaller disperion in FWHM ($\sim0.05\arcsec$) but slightly larger dispersion in ellipticity ($\sim0.25$). They also show the recovered astrometric errors for the two clusters as a function of magnitude, with NGC 3201 showing notably worse errors overall. Again, this is generally in agreement with our findings.

Although these results are preliminary and do not account for all the sources of error in the MAVIS astrometric error budget, they are encouraging as they account for the largest terms in the budget at this point. In the case of simulations II and III, representing the most realistic NGS constellations for the astrometric science cases, an astrometric accuracy of 50~$\mu$as is achieved in all magnitude bins brighter than $m=19$ (neglecting the saturated stars in the case of Simulation II) for an exposure time of 30s. This equates to about 20\% of the recovered stars. An astrometric accuracy of 150~$\mu$as is achieved in bins down to $m=22$ or about 50\% of the recovered stars. The results are also encouraging in that they demonstrate the successful application of current PSF-fitting photometric methods to simulated MAVIS data, in particular the idealised Fourier modeled PSF with its unique eight pronged pinwheel. 

\subsection{Detecting a Central IMBH in NGC 3201}
\begin{figure}
\includegraphics[width=\linewidth]{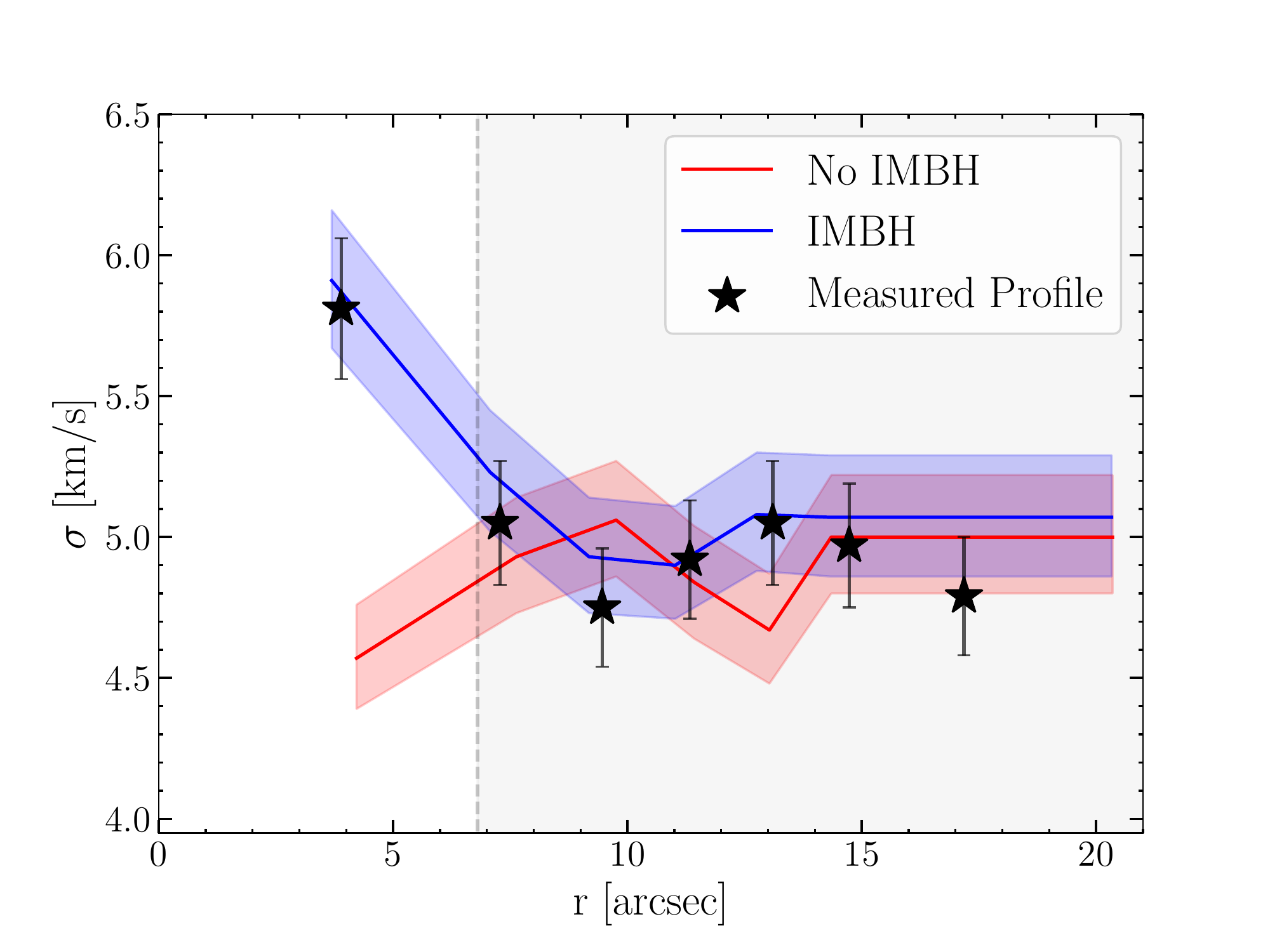}
\caption{Velocity dispersion profile recovered from simulated images over a ten year epoch. Black symbols represent the binned average for stars with a V band magnitude greater than 24. Each bin contains 300 stars (the outer-most bin contains on 297). The blue region denotes the input N-body model with the 1500 M$_{\odot}$ IMBH, using the same magnitude limit. The red region represents the N-body model without an IMBH. Vertical dashed line indicates the typical radial limit of HST proper motion measurements, which miss the crucial central regions where the IMBH signature is maximum.}
\label{fig:imbhresults}
\end{figure}

The results of the simulation described in Section~\ref{sec:imbh} are shown in Figs.~\ref{fig:pmresults} and \ref{fig:imbhresults} where we show the PM results and the recovered cluster velocity profile respectively. Beginning with Fig.~\ref{fig:pmresults}, we show the difference between the input and measured PMs over the ten year epoch as a function of magnitude (top panel) and cluster-centric distance (bottom panel). The matched catalogue has been filtered to include only the best matches (1228 matches kept of 1375 made), quantified by the matching distance recovered by the cKDTree. Individual errors attached to each recovered star are computed as as $\sigma_{\mathrm{PM X,Y}} = \sqrt{2}\frac{(\mu_{\mathrm{input}} - \mu_{\mathrm{meas.}})}{10}$. In both plots we also highlight the stars with recovered PM accuracies of 50~$\mu$as or better and 150~$\mu$as or better as the purple and indigo stars respectively.  Without filtering we find that 1182 of the 1194 recovered stars with magnitudes $m\leq25$ have a PM accuracy of 50~$\mu$arcsec or better, this number reduces to 97/181 stars for stars with magnitudes $m>25$. After filtering by matching distance, we find 1142/1143 stars with magnitudes $m\leq25$ and 73/85 with magnitude $m>25$ have a PM accuracy of 50~$\mu$arcsec or better. From the bottom panel, PM accuracy as a function of cluster-centric radius, we see no field dependence on the astrometric accuracy. This confirms the results of Section~\ref{sec:astrobudget}, that the HO PSF field variability \todo{(mainly anisoplanatism)} does not have a strong effect on the final astrometric performance. 

In Fig.~\ref{fig:imbhresults} we present a most exciting result, the recovery of the dynamical signature of the planted 1500~M$_{\odot}$ IMBH at the centre of the simulated cluster. The signature can be seen in the inner $\sim6$~\arcsec of the cluster, a region not accessible by HST given the crowding limit for a cluster at approximately the same distance \citep{hstpromo2}. The HST crowding limit is marked with a dashed line and shaded region. To create the dispersion profile shown as the black stars, we bin the catalogue in groups of 300 and determine the standard deviation of each bin to create the error bars shown in the figure. To compare the recovered profile with the theoretical IMBH and no-IMBH profiles, we create dispersion profiles directly from the input catalogues in the same way as the recovered data. These are shown as the blue and red profiles and shaded regions respectively. We find good agreement between the measured dispersion profile and the input IMBH dispersion profile. Comparing the recovered data with the no-IMBH profile, we see a clear departure in the inner $\sim7$~\arcsec.

Although comparing our dispersion profile against those  of NGC~3201 from the literature is not meaningful, we can compare error estimates at various radii to see the potential gain of MAVIS in the cluster centre. We first compare our results to the dispersion profile of NGC~3201 created by \cite{baumgardt19} using \textit{Gaia} DR2 data \cite{gaiadr2}. To do this, we take the dispersion values listed in their Table 4 for bins ranging in clustercentric distance from $r_{\mathrm{min}}=57.75\arcsec$ to $r_{\mathrm{max}}=384.09\arcsec$ and convert them to km/s assuming the same distance as \cite{baumgardt19} for the cluster (4.6~kpc). The resulting dispersion errors are fairly consistent across the radial bins, with an average upper limit of 0.26~km/s and lower limit of 0.24~km/s. These numbers are almost identical to the dispersion error associated with the \textit{inner-most} radial bin in our profile ($r_{\mathrm{min}}=3.89\arcsec$, $\pm0.23$~km/s). Note that we have a similar number of stars per bin as  \cite{baumgardt19} ($\sim300$). Our dispersion errors decrease as a function of increasing clustercentric radius, as one would expect, but do not vary drastically. The smallest dispersion error we find is $\pm0.20$~km/s in the furthest bin, $r_{\mathrm{max}}=17.2\arcsec$. 

At present there is no published dispersion profile for NGC~3201 using HST, however we can compare our results to a similar cluster studied in the HSTPROMO study \citep{hstpromo1,hstpromo2}. The MW GC NGC~6752 is similar in mass to NGC~3201, although it is slightly more massive and centrally concentrated \citep{baumgardt18}. Nonetheless, the HST cluster dispersion profile gives us a rough estimate of the expected performance of HST in NGC~3201. Using the distance listed in the catalogue of \cite{baumgardt18} we converted the PM dispersion profile of NGC~6752 published in \cite{hstpromo2} to km/s. A dispersion error of $\pm$0.81~km/s was measured in the inner $r_{\mathrm{min}}=2.74\arcsec$ of the cluster using 25 stars. If we resize our bins to include 150 stars each, we are able to probe the inner $2.73\arcsec$ of our synthetic NGC~3201-like cluster and recover a dispersion error of $\pm0.35$~km/s. Having significantly more stars in the inner radial bin naturally results in a smaller overall error, thus the more impressive takeaway is the factor six times more stars recoverable by MAVIS in same inner region of the cluster. 

We can also compare our dispersion profile to one generated using radial velocities measurements. To do this, we turn to the study of \cite{kamann18} using MUSE wide-field mode. Within the core radius, $r_{c}=1.3\arcmin$ \citep{baumgardt18} \cite{kamann18} compute an internal velocity dispersion with an associated uncertainty of $\pm0.5$~km/s by weighting $\sim27,000$ spectra of $\sim4000$ stars. They also determine the dispersion in smaller bin intervals ranging from $r_{\mathrm{min}}=5.8\arcsec$ to $r_{\mathrm{min}}=63.0\arcsec$, with at least 100 stars per bin and associated errors ranging from $\sim\pm0.44$~km/s in the inner-most bin to $0.21$~km/s in the outer-most bin. In the second-outer-most bin, presumably the most populated, at $r=46.5\arcsec$, they achieve an impressively small dispersion error of $\sim\pm0.10$~km/s. If we re-size our bins to include exactly 100 stars, we find a comparable dispersion error of $\pm0.45$~km/s in the inner $2.19\arcsec$ of the cluster. Therefore, we find our predicted astrometric accuracy to be comparable if not better beyond the crowding limit of MUSE in the seeing-limited mode. However, this improvement equalises with increasing clustercentric radius, in-line with decreased crowding and likely increased bin populations. 

A recent study to examine the internal kinematics of the MW GC M15 was made by \cite{usher2021} using MUSE in narrow field mode, the AO-assisted mode utilising the GALACSI AO module \citep{galacsi1, galacsi2}. Although M15 is more than twice as distant as NGC~3201 \citep{baumgardt18}, we can explore the gain in GC studies by MUSE with the addition of AO. \cite{usher2021} report an increase of a factor of $\sim5$ in the number of measurable stars within the inner 1~\arcsec of the cluster (33 vs. 7) and a factor $\sim2$ increase in the number of measurable stars within the inner $8~\arcsec$ (801 vs. 363). They also report an average FWHM of 0.08~$\arcsec$ for their AO-corrected PSFs at $H_{\alpha}$ (656.3~nm). Recall that MAVIS is expected to provide AO-corrected PSF FWHM of $0.0179\arcsec$ at 550~nm and thus we can expect to measure even more stars within the crowded cluster centres of distant MW GCs like M15.  

Bearing in mind that this simulation is likely optimistic, we have shown that MAVIS will be able to measure stellar velocities in the crucial central few arcseconds as accurately from proper motions as other instruments will at much larger radii from spectroscopic radial velocities. And we have shown that MAVIS has the potential to out-perform both \textit{Gaia} and HST in cluster centres. We can also conclude from this simulation that MAVIS has the potential to detect \textit{modestly} sized IMBHs in GC centres. Finally, a handful of high-velocity stars ($\sim30$~km/s) were also recovered in the centre of the MAVIS FoV, highlighting an additional potential capability of MAVIS, recovering a predicted population of high-velocity stars in the vicinity of an IMBH \citep{baumgardt19b}. Ultimately, this simulation provides an exciting glimpse into the astrometric capabilities of MAVIS, both for the indirect detection of IMBHs in GCs from thousands of accurate stellar proper motions, and for the direct detection of an IMBH through a handful of accurately measured high-velocity stars.

\section{Future Work}
\label{sec:futwork}
Although the results presented in the previous section are encouraging, they are preliminary. There are several outstanding sources of error, and methods to model expected errors, that must be tested to complete a comprehensive astrometric budget for MAVIS. In this section we discuss two additional sources of error, atmospheric dispersion and dynamic distortion, and discuss improvements to \texttt{MAVISIM} including; moving to an end-to-end PSF and modeling the calibration procedure for the imager. Additionally, though discussed only briefly, we would like to further investigate the use of an alternative analysis tool to \texttt{DAOPhot}.

\subsection{Dynamic Distortion Term}
Unlike the LO static distortion discussed in Sec.~\ref{sec:lowordertheory}, which can be calibrated without on-sky measurements, dynamic distortion must be calibrated on-sky using astrometric reference stars. Sources of dynamic distortion include; gravitational and thermal flexure of the entire telescope-instrument optical relay. In the case of GeMS, \cite{lu14} found that dynamic distortions contributed a 2000$\mu$as LO RMS error and 600$\mu$as HO RMS error (after fitting) introduced over two consecutive nights of observing. This is compared to the 790$\mu$as LO RMS error and 460$\mu$as HO RMS error measured in a single night. Despite fitting with a high order polynomial, they found residual distortions in the case of multi-epoch astrometry. \cite{lu14} identified the source of dynamic distortion as the flexure induced by an elevation change of 10$^{\mathrm{o}}$. Dynamic distortions from gravitational flexure have proven to be the dominant terms in the GeMS astrometric error budget \citep{neichel14}. Because MAVIS will be located on the Nasmyth platform of the VLT it will be significantly more resilient to gravitationally induced flexure. Additionally, much work is being done to deliver deliver stable temperature controls.

In this first iteration of \texttt{MAVISIM}, we do not attempt to model dynamic distortions. This decision is made to avoid any oversimplification of dynamic terms through modeling as simple perturbations of the static distortion map. In the next iteration we will explore a method similar to that of \cite{micadodist}. That is, realistic perturbations are introduced as the misalignment of different optical elements to simulate gravitational and thermal flexures. Ray tracing techniques will then be used to generate the final distortion mask for each set of perturbations, this is then done for several trials in a Monte Carlo-style simulation. 

\subsection{Atmospheric Dispersion}

Throughout this first study using \texttt{MAVISIM} we adopt a monochromatic PSF at the wavelength of optimised performance for MAVIS, 550~nm. In the next iteration of the tool we will use broadband PSFs to better investigate the influence of any chromatic terms in the astrometry budget, with the most obvious term originating from atmospheric refraction. As light traverses the atmosphere it experiences refraction that is both a function of the incident angle and wavelength of the light. In the broadband situation this results in chromatic dispersion that elongates the PSF shape. \cite{bornadc} recently investigated the effect of chromatic dispersion on the astrometic budget of MICADO. They examined the effects of input atmospheric models, the chromatic dispersion internal to the MICADO optical design and the predicted capabilities of the atmospheric dispersion corrector. They found that the optical design dominated the dispersion budget. As such, we will investigate the contribution from the MAVIS optomechanical design using an end-to-end approach with Zemax. The spatial dispersion of the PSF will be calculated at several wavelengths and a broadband PSF will be then created from a weighted average of the individual PSFs.

\subsection{\label{sec:calibmask} Calibrating the MAVIS Imager}
Designing an MCAO system with highly accurate astrometry (50-150~$\mu$as) as a goal requires a combined approach to both minimise the internal static distortion, and model the remaining distortion well such that it can be calibrated out \citep{micadodist}. Static distortion, of the kind described in Sec.~\ref{sec:lowordertheory}, can be calibrated using an astrometric mask, although to what precision is unclear. Both NFIRAOS/IRIS \citep[InfraRed  Imaging  Spectrograph][]{iris} and MAORY/MICADO have astrometric calibration masks planned as part of their calibration units \citep[e.g. NFIRAOS/IRIS, MAORY/MICADO:][]{tmtmask1, tmtmask2, micadomask}. 

The MAVIS team is undertaking an investigation into the ideal calibration mask pattern and sampling to recover the static distortion pattern within the AOM. As part of this experimentation we are investigating the method by which the distortion pattern is recovered. As direct calibration using an absolute reference frame may prove difficult (e.g. manufacturing difficulties), we are investigating using a ``self-calibration method''. This technique involves capturing many dithered images and then recovering the distortion pattern through measurements of the gradient of the positions of pinholes between dithered frames \citep{tmta04elt5}. We will use the static distortion modeling capabilities of \texttt{MAVISIM} to investigate this.

\begin{figure}
\includegraphics[width=\linewidth]{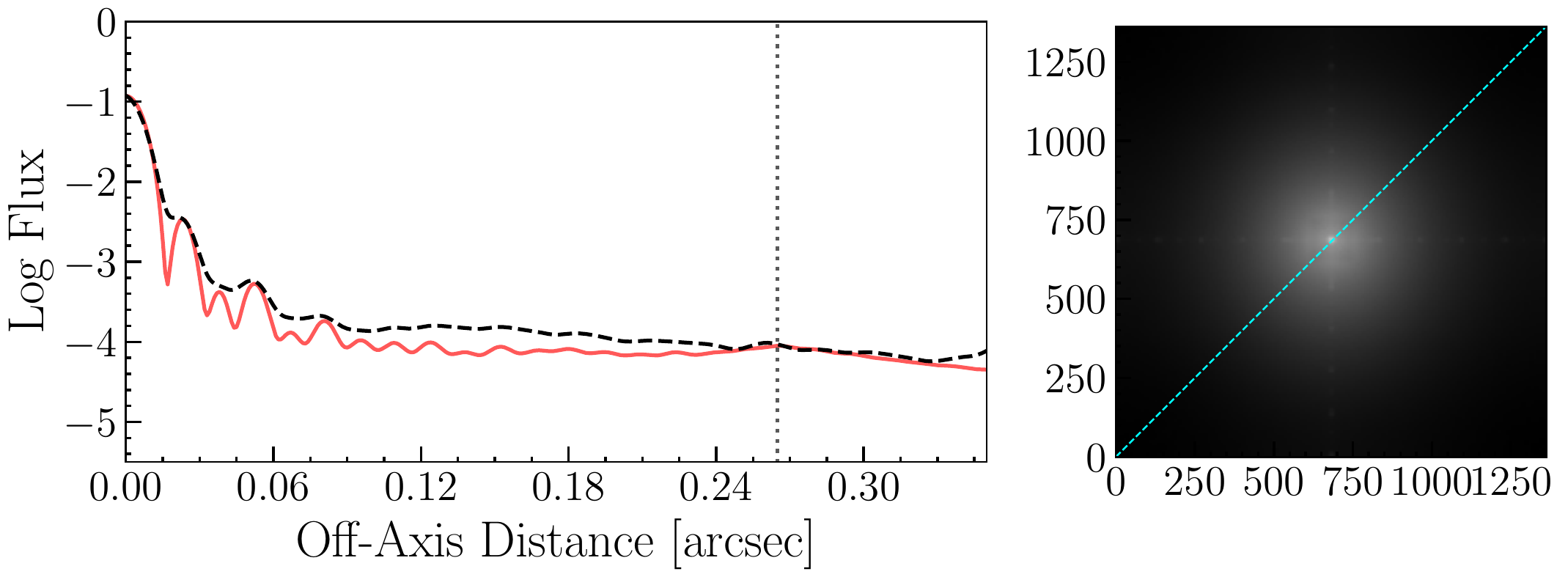}
\caption{
Normalised radial profile (dashed line) of the end-to-end MAVIS PSF generated in \textsc{COMPASS} sampled in the centre of the MAVIS FoV. The Fourier PSF sampled in the centre of the FoV is plotted beneath the end-to-end profile in red, with the theoretical control radius shown as the dotted gray line. Note the difference in distribution of power between the PSF models. A thumbnail of the end-to-end log intensity of the PSF is given on the right.}
\label{fig:eteprofile}
\end{figure}

\subsection{\label{sec:etepsf} \texttt{MAVISIM} 1.1: Moving to an End-to-End Spatially Variable PSF}

As discussed in Section~\ref{sec:psfvarmodel}, an end-to-end (e2e) PSF model has been developed for MAVIS in \textsc{COMPASS} \citep{compass}. The details of which can be found in \cite{jessemavis}. Recall that the rapid adjustment of the input NGS constellation and thus exploration of the TT residual is not possible using an e2e PSF and therefore it was not appropriate for this study. However, \todo{as we explore science cases that require both accurate astrometry \textit{and} photometry like MW GCs, we will move to using the e2e PSF to investigate both. As in the case of NGC~3201, most MW GCs are likely to contain NGS distributed most like the ``good'' NGS constellation and thus we can fix the the input TT contribution.} The e2e PSF is more appropriate for a study involving photometry study as it presents a more realistic representation of the PSF beyond the core. This is shown in Fig.~\ref{fig:eteprofile}, where the radial profile of the central e2e PSF is compared to the profile of the central Fourier PSF. The theoretical control radius is marked with a gray dashed line and is clearly distinguishable in the case of the Fourier PSF. This is not the case of the e2e PSF. Similar techniques will be used to model the PSF for photometric studies but more attention will be paid to accurately modeling the PSF out to larger radii.

\section{Conclusions}
We have introduced the novel image simulating tool, \texttt{MAVISIM}, designed to model the major sources of astrometric error in multi-conjugate adaptive optics (MCAO) systems, with a specific application to The MCAO Visible Imager and Spectrograph (MAVIS) instrument. In the current iteration of \texttt{MAVISIM} we model i) high (HO) and low order (LO) point spread function (PSF) field variability, ii) a tip-tilt (TT)  residual error, iii) static field distortion and iv) telescope and adaptive optics module throughput and detector characteristics.  In this study we have used \texttt{MAVISIM} to examine, for the first time, the expected astrometric accuracy of the system after accounting for these terms. \todo{Note that these errors represent only a small subset of the error terms present in our preliminary astrometric budget (Table~\ref{tab:astrobudget}) and thus we are investigating an incomplete and surely optimistic picture of the astrometric abilities of MAVIS.} To derive an initial astrometric accuracy, we have generated eight different images using \texttt{MAVISIM} to investigate the effects individually and in tandem. We have analysed these images using the standard astrometric analysis tool, \texttt{DAOPhot-IV} \citep{daophot1, daophot2} to perform PSF-fitting photometry and extract stellar positions. 

The first result of the simulations is that the LO TT residual error is more important at faint magnitudes than the HO PSF field variability. The TT residual term is associated with imperfect corrections to the LO AO distortion terms and is a direct reflection of the natural guide star (NGS) constellation geometry and NGS magnitudes. We find that HO PSF field variability associated with the laser guide star (LGS) constellation geometry has little effect on the simulated astrometric capabilities when performing PSF-fitting photometry using a spatially variable model of the PSF. 

The second result of the simulations is that the \textit{magnitude} of the TT residual term is more important than the field variability. This conclusion was reached through performing three simulations using different NGS constellations with associated TT residual maps. Each constellation was comprised of slightly different NGS geometry and magnitudes. The worst constellation, comprised of the faintest NGSs located the furthest from the centre of the FoV had the largest overall, but smoothest varying, TT residual map. When compared against the astrometric accuracies associated with a typical NGS constellation, showing the highest spatial variability in the TT residual, the worst constellation yielded the worst results. This result was found to be in agreement with existing literature studies of Milky Way (MW) globular clusters (GCs), though a direct comparison was not possible.

The final result from the simulations show that MAVIS \todo{\textit{could}} achieve an astrometric accuracy of 50~$\mu$as for stars with magnitude $m\leq19$ in a 30s exposure. \todo{Bearing in mind that we have yet to model all the astrometric error terms in the MAVIS astrometric budget as discussed in Section~\ref{sec:futwork} and thus these results are optimistic. Nevertheless,} this accuracy was achieved in the case of both the best and typical NGS constellations, with the best constellation likely to be found in stellar fields, like GCs, and the typical constellation being more general. Using the same NGS configurations, an astrometric accuracy of 150~$\mu$as was measured for stars with magnitude $m=22$, equating to about 50\% of the total number of measured stars.

To test the simulated capabilities of MAVIS on a real science case, we used \texttt{MAVISIM} to investigate the likelihood of dynamically detecting an intermediate mass black hole (IMBH) at the centre of a MW GC. We used two N-body catalogues, specifically modeled after the MW GC NGC~3201 as input and simulated a 10 year epoch accounting for all astrometric effects, while assuming the best-case NGS constellation. One of the catalogues contained the signature of a central 1500 M$_{\odot}$ IMBH, while the other did not. The recovered PM accuracy in the catalogue containing the IMBH was found to be within 50~$\mu$arcsec for 97/181 of the recovered stars with magnitudes $m\geq25$ and for 1182/1194 of the recovered stars with magnitudes $m\leq25$. We also confirm in our PM study that the LO TT residual term is the dominant source of astrometric error, not HO PSF field variability, as the PM accuracies show no obvious field dependence. Comparing our recovered velocity dispersion profiles between the two catalogues, we show that MAVIS \todo{\textit{could} under favourable astrometric conditions,} recover the dynamical signature of the IMBH and that the signature occurs within the typical crowding limit of HST. Finally, comparing our velocity dispersion errors with profiles of NGC~3201 made using \textit{Gaia} (DR2) and MUSE data, we find that MAVIS could deliver dispersion errors in the inner $\sim4\arcsec$ of the cluster with the same precision as both \textit{Gaia} and MUSE deliver in the outer$\sim60\arcsec$ of the cluster. This demonstrates the potential strength of MAVIS in the extremely crowded regime, paving the way for exciting new science.

This study is only the beginning of a much larger investigation into the astrometric capabilities of the next generation ESO instrument MAVIS. In the next iteration of the tool we will investigate the effects of dynamical distortion, adopt a more realistic end-to-end model of the MAVIS PSF, investigate atmospheric effects in more detail and experiment with calibration of the expected astrometric distortion. Ultimately, the astrometric error budget will need to both; account for as many error terms as can be modeled or calculated accurately, and feed back directly into the core MAVIS science cases. \texttt{MAVISIM} provides the means to do both and the intention is to improve and test the tool as frequently as possible until MAVIS goes on-sky. Once on-sky, MAVIS will open up a hitherto unexplored regime of wide field AO in the visible, serving as a workhorse instrument with a broad and well balanced science portfolio. 

\section{Data Availability} 
The data underlying this article are available in the \textit{example} directory on GitHub at \url{https://github.com/smonty93/MAVISIM}. \texttt{MAVISIM} is also available through this repository.

\section*{Acknowledgements}
We thank the anonymous referee very much for their helpful comments that greatly improved the paper. This work relies heavily on the Astropy \citep{astropy1, astropy2}, SciPy \citep{scipy}, NumPy \citep{numpy} and Matplotlib \citep{matplotlib} libraries. SM wishes to thank the entire MAVIS team of wonderful collaborators for many fruitful discussions, support and encouragement. She also wishes to acknowledge the traditional custodians of Mt. Stromlo, the Ngunawal people and pay her respect to elders past and present. SM acknowledges funding support from the the Natural Sciences and Engineering Research Council of Canada (NSERC), [funding reference number PGSD3 - 545852 - 2020]. Cette recherche a été financée par le Conseil de recherches en sciences naturelles et en génie du Canada (CRSNG), [numéro de référence PGSD3 - 545852 - 2020].




\bibliographystyle{mnras}
\bibliography{bib} 

\begin{thebibliography}{}
\makeatletter
\relax
\def\mn@urlcharsother{\let\do\@makeother \do\$\do\&\do\#\do\^\do\_\do\%\do\~}
\def\mn@doi{\begingroup\mn@urlcharsother \@ifnextchar [ {\mn@doi@}
  {\mn@doi@[]}}
\def\mn@doi@[#1]#2{\def\@tempa{#1}\ifx\@tempa\@empty \href
  {http://dx.doi.org/#2} {doi:#2}\else \href {http://dx.doi.org/#2} {#1}\fi
  \endgroup}
\def\mn@eprint#1#2{\mn@eprint@#1:#2::\@nil}
\def\mn@eprint@arXiv#1{\href {http://arxiv.org/abs/#1} {{\tt arXiv:#1}}}
\def\mn@eprint@dblp#1{\href {http://dblp.uni-trier.de/rec/bibtex/#1.xml}
  {dblp:#1}}
\def\mn@eprint@#1:#2:#3:#4\@nil{\def\@tempa {#1}\def\@tempb {#2}\def\@tempc
  {#3}\ifx \@tempc \@empty \let \@tempc \@tempb \let \@tempb \@tempa \fi \ifx
  \@tempb \@empty \def\@tempb {arXiv}\fi \@ifundefined
  {mn@eprint@\@tempb}{\@tempb:\@tempc}{\expandafter \expandafter \csname
  mn@eprint@\@tempb\endcsname \expandafter{\@tempc}}}

\bibitem[\protect\citeauthoryear{{Agapito}, {Vassallo}, {Plantet}, {Viotto},
  {Pinna}, {Neichel}, {Fusco}  \& {Rigaut}}{{Agapito}
  et~al.}{2020}]{agapito2020}
{Agapito} G.,  {Vassallo} D.,  {Plantet} C.,  {Viotto} V.,  {Pinna} E.,
  {Neichel} B.,  {Fusco} T.,   {Rigaut} F.,  2020, in Society of Photo-Optical
  Instrumentation Engineers (SPIE) Conference Series. p. 114483R (\mn@eprint
  {arXiv} {2012.14487}), \mn@doi{10.1117/12.2561252}

\bibitem[\protect\citeauthoryear{{Anderson} \& {King}}{{Anderson} \&
  {King}}{2003}]{anderson03}
{Anderson} J.,  {King} I.~R.,  2003, \mn@doi [\pasp] {10.1086/345491}, \href
  {https://ui.adsabs.harvard.edu/abs/2003PASP..115..113A} {115, 113}

\bibitem[\protect\citeauthoryear{{Anderson} \& {van der Marel}}{{Anderson} \&
  {van der Marel}}{2010}]{anderson10}
{Anderson} J.,  {van der Marel} R.~P.,  2010, \mn@doi [\apj]
  {10.1088/0004-637X/710/2/1032}, \href
  {https://ui.adsabs.harvard.edu/abs/2010ApJ...710.1032A} {710, 1032}

\bibitem[\protect\citeauthoryear{{Anderson}, {King}  \& {Meylan}}{{Anderson}
  et~al.}{1998}]{anderson98}
{Anderson} J.,  {King} I.~R.,   {Meylan} G.,  1998, in American Astronomical
  Society Meeting Abstracts. p. 68.02

\bibitem[\protect\citeauthoryear{{Arsenault} et~al.,}{{Arsenault}
  et~al.}{2012}]{galacsi1}
{Arsenault} R.,  et~al., 2012, in {Ellerbroek} B.~L.,  {Marchetti} E.,
  {V{\'e}ran} J.-P.,  eds,  Society of Photo-Optical Instrumentation Engineers
  (SPIE) Conference Series Vol. 8447, Adaptive Optics Systems III. p. 84470J,
  \mn@doi{10.1117/12.926074}

\bibitem[\protect\citeauthoryear{{Astropy Collaboration} et~al.,}{{Astropy
  Collaboration} et~al.}{2013}]{astropy1}
{Astropy Collaboration} et~al., 2013, \mn@doi [\aap]
  {10.1051/0004-6361/201322068}, \href
  {https://ui.adsabs.harvard.edu/abs/2013A&A...558A..33A} {558, A33}

\bibitem[\protect\citeauthoryear{{Astropy Collaboration} et~al.,}{{Astropy
  Collaboration} et~al.}{2018}]{astropy2}
{Astropy Collaboration} et~al., 2018, \mn@doi [\aj] {10.3847/1538-3881/aabc4f},
  \href {https://ui.adsabs.harvard.edu/abs/2018AJ....156..123A} {156, 123}

\bibitem[\protect\citeauthoryear{{Bacon} et~al.,}{{Bacon} et~al.}{2010}]{muse}
{Bacon} R.,  et~al., 2010, {The MUSE second-generation VLT instrument}.
p. 773508, \mn@doi{10.1117/12.856027}

\bibitem[\protect\citeauthoryear{{Baumgardt}}{{Baumgardt}}{2017}]{baumgardt17}
{Baumgardt} H.,  2017, \mn@doi [\mnras] {10.1093/mnras/stw2488}, \href
  {https://ui.adsabs.harvard.edu/abs/2017MNRAS.464.2174B} {464, 2174}

\bibitem[\protect\citeauthoryear{{Baumgardt} \& {Hilker}}{{Baumgardt} \&
  {Hilker}}{2018}]{baumgardt18}
{Baumgardt} H.,  {Hilker} M.,  2018, \mn@doi [\mnras] {10.1093/mnras/sty1057},
  \href {https://ui.adsabs.harvard.edu/abs/2018MNRAS.478.1520B} {478, 1520}

\bibitem[\protect\citeauthoryear{{Baumgardt}, {Hilker}, {Sollima}  \&
  {Bellini}}{{Baumgardt} et~al.}{2019a}]{baumgardt19}
{Baumgardt} H.,  {Hilker} M.,  {Sollima} A.,   {Bellini} A.,  2019a, \mn@doi
  [\mnras] {10.1093/mnras/sty2997}, \href
  {https://ui.adsabs.harvard.edu/abs/2019MNRAS.482.5138B} {482, 5138}

\bibitem[\protect\citeauthoryear{{Baumgardt} et~al.,}{{Baumgardt}
  et~al.}{2019b}]{baumgardt19b}
{Baumgardt} H.,  et~al., 2019b, \mn@doi [\mnras] {10.1093/mnras/stz2060}, \href
  {https://ui.adsabs.harvard.edu/abs/2019MNRAS.488.5340B} {488, 5340}

\bibitem[\protect\citeauthoryear{{Beckers}}{{Beckers}}{1988}]{beckers1988}
{Beckers} J.~M.,  1988, in Very Large Telescopes and their Instrumentation,
  Vol. 2. pp 693--703

\bibitem[\protect\citeauthoryear{{Bellini} et~al.,}{{Bellini}
  et~al.}{2014}]{hstpromo1}
{Bellini} A.,  et~al., 2014, \mn@doi [\apj] {10.1088/0004-637X/797/2/115},
  \href {https://ui.adsabs.harvard.edu/abs/2014ApJ...797..115B} {797, 115}

\bibitem[\protect\citeauthoryear{{Bellini}, {Bianchini}, {Varri}, {Anderson},
  {Piotto}, {van der Marel}, {Vesperini}  \& {Watkins}}{{Bellini}
  et~al.}{2017}]{bellini17}
{Bellini} A.,  {Bianchini} P.,  {Varri} A.~L.,  {Anderson} J.,  {Piotto} G.,
  {van der Marel} R.~P.,  {Vesperini} E.,   {Watkins} L.~L.,  2017, \mn@doi
  [\apj] {10.3847/1538-4357/aa7c5f}, \href
  {https://ui.adsabs.harvard.edu/abs/2017ApJ...844..167B} {844, 167}

\bibitem[\protect\citeauthoryear{{Beltramo-Martin}, {Marasco}, {Fusco},
  {Massari}, {Milli}, {Fiorentino}  \& {Neichel}}{{Beltramo-Martin}
  et~al.}{2020}]{beltramo2020}
{Beltramo-Martin} O.,  {Marasco} A.,  {Fusco} T.,  {Massari} D.,  {Milli} J.,
  {Fiorentino} G.,   {Neichel} B.,  2020, \mn@doi [\mnras]
  {10.1093/mnras/staa525}, \href
  {https://ui.adsabs.harvard.edu/abs/2020MNRAS.494..775B} {494, 775}

\bibitem[\protect\citeauthoryear{{Cameron}, {Britton}  \& {Kulkarni}}{{Cameron}
  et~al.}{2009}]{cameron09}
{Cameron} P.~B.,  {Britton} M.~C.,   {Kulkarni} S.~R.,  2009, \mn@doi [\aj]
  {10.1088/0004-6256/137/1/83}, \href
  {https://ui.adsabs.harvard.edu/abs/2009AJ....137...83C} {137, 83}

\bibitem[\protect\citeauthoryear{{Crane} et~al.,}{{Crane}
  et~al.}{2018}]{nfiraos2018}
{Crane} J.,  et~al., 2018, in {Close} L.~M.,  {Schreiber} L.,   {Schmidt} D.,
  eds,  Society of Photo-Optical Instrumentation Engineers (SPIE) Conference
  Series Vol. 10703, Adaptive Optics Systems VI. p. 107033V,
  \mn@doi{10.1117/12.2314341}

\bibitem[\protect\citeauthoryear{Cranney, Don{\'a}, Rigaut  \&
  Korkiakoski}{Cranney et~al.}{2019}]{jessemavis}
Cranney J.,  Don{\'a} J.~D.,  Rigaut F.,   Korkiakoski V.,  2019, in
  Proceedings of the 6th AO4ELT Conference. \url
  {http://ao4elt6.copl.ulaval.ca/proceedings/401-qDC4-251.pdf}

\bibitem[\protect\citeauthoryear{{Dalessandro} et~al.,}{{Dalessandro}
  et~al.}{2016}]{dalessandro16}
{Dalessandro} E.,  et~al., 2016, \mn@doi [\apj] {10.3847/1538-4357/833/1/111},
  \href {https://ui.adsabs.harvard.edu/abs/2016ApJ...833..111D} {833, 111}

\bibitem[\protect\citeauthoryear{{Davies} et~al.,}{{Davies}
  et~al.}{2016}]{micado}
{Davies} R.,  et~al., 2016, {MICADO: first light imager for the E-ELT}.
p. 99081Z, \mn@doi{10.1117/12.2233047}

\bibitem[\protect\citeauthoryear{{Dicke}}{{Dicke}}{1975}]{dicke1975}
{Dicke} R.~H.,  1975, \mn@doi [\apj] {10.1086/153639}, \href
  {https://ui.adsabs.harvard.edu/abs/1975ApJ...198..605D} {198, 605}

\bibitem[\protect\citeauthoryear{{Ebisuzaki} et~al.,}{{Ebisuzaki}
  et~al.}{2001}]{ebisuzaki01}
{Ebisuzaki} T.,  et~al., 2001, \mn@doi [\apjl] {10.1086/338118}, \href
  {https://ui.adsabs.harvard.edu/abs/2001ApJ...562L..19E} {562, L19}

\bibitem[\protect\citeauthoryear{{Ellerbroek}}{{Ellerbroek}}{1994}]{ellerbroek1994}
{Ellerbroek} B.~L.,  1994, \mn@doi [Journal of the Optical Society of America
  A] {10.1364/JOSAA.11.000783}, \href
  {https://ui.adsabs.harvard.edu/abs/1994JOSAA..11..783E} {11, 783}

\bibitem[\protect\citeauthoryear{{Esposito} et~al.,}{{Esposito}
  et~al.}{2010}]{lbtao}
{Esposito} S.,  et~al., 2010, in {Ellerbroek} B.~L.,  {Hart} M.,  {Hubin} N.,
  {Wizinowich} P.~L.,  eds,  Society of Photo-Optical Instrumentation Engineers
  (SPIE) Conference Series Vol. 7736, Adaptive Optics Systems II. p. 773609,
  \mn@doi{10.1117/12.858194}

\bibitem[\protect\citeauthoryear{{Ferrarese} \& {Ford}}{{Ferrarese} \&
  {Ford}}{2005}]{ferrarese05}
{Ferrarese} L.,  {Ford} H.,  2005, \mn@doi [\ssr] {10.1007/s11214-005-3947-6},
  \href {https://ui.adsabs.harvard.edu/abs/2005SSRv..116..523F} {116, 523}

\bibitem[\protect\citeauthoryear{{Fiorentino} et~al.,}{{Fiorentino}
  et~al.}{2017}]{micadoscience}
{Fiorentino} G.,  et~al., 2017, arXiv e-prints, \href
  {https://ui.adsabs.harvard.edu/abs/2017arXiv171204222F} {p. arXiv:1712.04222}

\bibitem[\protect\citeauthoryear{{Fiorentino} et~al.,}{{Fiorentino}
  et~al.}{2020}]{fiorentino20}
{Fiorentino} G.,  et~al., 2020, \mn@doi [\mnras] {10.1093/mnras/staa869}, \href
  {https://ui.adsabs.harvard.edu/abs/2020MNRAS.tmp.1061F} {}

\bibitem[\protect\citeauthoryear{{Freire} et~al.,}{{Freire}
  et~al.}{2017}]{freire17}
{Freire} P.~C.~C.,  et~al., 2017, \mn@doi [\mnras] {10.1093/mnras/stx1533},
  \href {https://ui.adsabs.harvard.edu/abs/2017MNRAS.471..857F} {471, 857}

\bibitem[\protect\citeauthoryear{Fried}{Fried}{1982}]{fried82}
Fried D.~L.,  1982, \mn@doi [J. Opt. Soc. Am.] {10.1364/JOSA.72.000052}, 72, 52

\bibitem[\protect\citeauthoryear{{Fusco} et~al.,}{{Fusco}
  et~al.}{2020}]{fusco2020}
{Fusco} T.,  et~al., 2020, \mn@doi [\aap] {10.1051/0004-6361/202037595}, \href
  {https://ui.adsabs.harvard.edu/abs/2020A&A...635A.208F} {635, A208}

\bibitem[\protect\citeauthoryear{{Gaia Collaboration} et~al.,}{{Gaia
  Collaboration} et~al.}{2016}]{gaiadr1}
{Gaia Collaboration} et~al., 2016, \mn@doi [\aap]
  {10.1051/0004-6361/201629512}, \href
  {https://ui.adsabs.harvard.edu/abs/2016A&A...595A...2G} {595, A2}

\bibitem[\protect\citeauthoryear{{Gaia Collaboration} et~al.,}{{Gaia
  Collaboration} et~al.}{2018}]{gaiadr2}
{Gaia Collaboration} et~al., 2018, \mn@doi [\aap]
  {10.1051/0004-6361/201833051}, \href
  {https://ui.adsabs.harvard.edu/abs/2018A&A...616A...1G} {616, A1}

\bibitem[\protect\citeauthoryear{{Gaia Collaboration}, {Brown}, {Vallenari},
  {Prusti}, {de Bruijne}, {Babusiaux}  \& {Biermann}}{{Gaia Collaboration}
  et~al.}{2020}]{gaiadr3}
{Gaia Collaboration} {Brown} A.~G.~A.,  {Vallenari} A.,  {Prusti} T.,  {de
  Bruijne} J.~H.~J.,  {Babusiaux} C.,   {Biermann} M.,  2020, arXiv e-prints,
  \href {https://ui.adsabs.harvard.edu/abs/2020arXiv201201533G} {p.
  arXiv:2012.01533}

\bibitem[\protect\citeauthoryear{Gebhardt, Rich  \& Ho}{Gebhardt
  et~al.}{2005}]{gebhardt05}
Gebhardt K.,  Rich R.~M.,   Ho L.~C.,  2005, \mn@doi [The Astrophysical
  Journal] {10.1086/497023}, 634, 1093

\bibitem[\protect\citeauthoryear{{Gendron} \& {Rousset}}{{Gendron} \&
  {Rousset}}{2012}]{gendron2012}
{Gendron} E.,  {Rousset} G.,  2012, in {Ellerbroek} B.~L.,  {Marchetti} E.,
  {V{\'e}ran} J.-P.,  eds,  Society of Photo-Optical Instrumentation Engineers
  (SPIE) Conference Series Vol. 8447, Adaptive Optics Systems III. p. 84476N,
  \mn@doi{10.1117/12.925554}

\bibitem[\protect\citeauthoryear{{Giersz}, {Leigh}, {Hypki}, {L{\"u}tzgendorf}
  \& {Askar}}{{Giersz} et~al.}{2015}]{giersz15}
{Giersz} M.,  {Leigh} N.,  {Hypki} A.,  {L{\"u}tzgendorf} N.,   {Askar} A.,
  2015, \mn@doi [\mnras] {10.1093/mnras/stv2162}, \href
  {https://ui.adsabs.harvard.edu/abs/2015MNRAS.454.3150G} {454, 3150}

\bibitem[\protect\citeauthoryear{{Gratadour} et~al.,}{{Gratadour}
  et~al.}{2014}]{compass}
{Gratadour} D.,  et~al., 2014, in {Marchetti} E.,  {Close} L.~M.,   {Veran}
  J.-P.,  eds,  Society of Photo-Optical Instrumentation Engineers (SPIE)
  Conference Series Vol. 9148, Adaptive Optics Systems IV. p. 91486O,
  \mn@doi{10.1117/12.2056358}

\bibitem[\protect\citeauthoryear{{Grindlay}, {Heinke}, {Edmonds}  \&
  {Murray}}{{Grindlay} et~al.}{2001}]{grindlay01}
{Grindlay} J.~E.,  {Heinke} C.,  {Edmonds} P.~D.,   {Murray} S.~S.,  2001,
  \mn@doi [Science] {10.1126/science.1061135}, \href
  {https://ui.adsabs.harvard.edu/abs/2001Sci...292.2290G} {292, 2290}

\bibitem[\protect\citeauthoryear{{G{\"u}rkan}, {Freitag}  \&
  {Rasio}}{{G{\"u}rkan} et~al.}{2004}]{gurken04}
{G{\"u}rkan} M.~A.,  {Freitag} M.,   {Rasio} F.~A.,  2004, \mn@doi [\apj]
  {10.1086/381968}, \href
  {https://ui.adsabs.harvard.edu/abs/2004ApJ...604..632G} {604, 632}

\bibitem[\protect\citeauthoryear{{H{\"a}berle} et~al.,}{{H{\"a}berle}
  et~al.}{2021}]{haberle2021}
{H{\"a}berle} M.,  et~al., 2021, arXiv e-prints, \href
  {https://ui.adsabs.harvard.edu/abs/2021arXiv210207782H} {p. arXiv:2102.07782}

\bibitem[\protect\citeauthoryear{Harris et~al.,}{Harris et~al.}{2020}]{numpy}
Harris C.~R.,  et~al., 2020, \mn@doi [Nature] {10.1038/s41586-020-2649-2}, 585,
  357

\bibitem[\protect\citeauthoryear{{Hayano} et~al.,}{{Hayano}
  et~al.}{2010}]{subaruao}
{Hayano} Y.,  et~al., 2010, in {Ellerbroek} B.~L.,  {Hart} M.,  {Hubin} N.,
  {Wizinowich} P.~L.,  eds,  Society of Photo-Optical Instrumentation Engineers
  (SPIE) Conference Series Vol. 7736, Adaptive Optics Systems II. p. 77360N,
  \mn@doi{10.1117/12.857567}

\bibitem[\protect\citeauthoryear{{H{\'e}nault-Brunet}, {Gieles}, {Strader},
  {Peuten}, {Balbinot}  \& {Douglas}}{{H{\'e}nault-Brunet}
  et~al.}{2020}]{brunet20}
{H{\'e}nault-Brunet} V.,  {Gieles} M.,  {Strader} J.,  {Peuten} M.,  {Balbinot}
  E.,   {Douglas} K.~E.~K.,  2020, \mn@doi [\mnras] {10.1093/mnras/stz2995},
  \href {https://ui.adsabs.harvard.edu/abs/2020MNRAS.491..113H} {491, 113}

\bibitem[\protect\citeauthoryear{{Herriot} et~al.,}{{Herriot}
  et~al.}{2000}]{altair}
{Herriot} G.,  et~al., 2000, in {Wizinowich} P.~L.,  ed.,  Society of
  Photo-Optical Instrumentation Engineers (SPIE) Conference Series Vol. 4007,
  Adaptive Optical Systems Technology. pp 115--125, \mn@doi{10.1117/12.390288}

\bibitem[\protect\citeauthoryear{{Herriot} et~al.,}{{Herriot}
  et~al.}{2010}]{herriot10}
{Herriot} G.,  et~al., 2010, in \procspie. p. 77360B,
  \mn@doi{10.1117/12.857662}

\bibitem[\protect\citeauthoryear{Hunter}{Hunter}{2007}]{matplotlib}
Hunter J.~D.,  2007, \mn@doi [Computing in Science \& Engineering]
  {10.1109/MCSE.2007.55}, 9, 90

\bibitem[\protect\citeauthoryear{{James}, {Boyer}, {Buchroeder}, {Ellerbroek}
  \& {Hunten}}{{James} et~al.}{2003}]{james03}
{James} E.,  {Boyer} C.,  {Buchroeder} R.~A.,  {Ellerbroek} B.~L.,   {Hunten}
  M.~R.,  2003, in {Wizinowich} P.~L.,  {Bonaccini} D.,  eds,  Society of
  Photo-Optical Instrumentation Engineers (SPIE) Conference Series Vol. 4839,
  \procspie. pp 67--80, \mn@doi{10.1117/12.457082}

\bibitem[\protect\citeauthoryear{{Jolissaint}, {V{\'e}ran}  \&
  {Conan}}{{Jolissaint} et~al.}{2006}]{jolissaint}
{Jolissaint} L.,  {V{\'e}ran} J.-P.,   {Conan} R.,  2006, \mn@doi [Journal of
  the Optical Society of America A] {10.1364/JOSAA.23.000382}, \href
  {https://ui.adsabs.harvard.edu/abs/2006JOSAA..23..382J} {23, 382}

\bibitem[\protect\citeauthoryear{{Kamann}, {Wisotzki}  \& {Roth}}{{Kamann}
  et~al.}{2013}]{kamann13}
{Kamann} S.,  {Wisotzki} L.,   {Roth} M.~M.,  2013, \mn@doi [\aap]
  {10.1051/0004-6361/201220476}, \href
  {https://ui.adsabs.harvard.edu/abs/2013A&A...549A..71K} {549, A71}

\bibitem[\protect\citeauthoryear{{Kamann} et~al.,}{{Kamann}
  et~al.}{2016}]{kamann16}
{Kamann} S.,  et~al., 2016, \mn@doi [\aap] {10.1051/0004-6361/201527065}, \href
  {https://ui.adsabs.harvard.edu/abs/2016A&A...588A.149K} {588, A149}

\bibitem[\protect\citeauthoryear{{Kamann} et~al.,}{{Kamann}
  et~al.}{2018}]{kamann18}
{Kamann} S.,  et~al., 2018, \mn@doi [\mnras] {10.1093/mnras/stx2719}, \href
  {https://ui.adsabs.harvard.edu/abs/2018MNRAS.473.5591K} {473, 5591}

\bibitem[\protect\citeauthoryear{{K{\i}z{\i}ltan}, {Baumgardt}  \&
  {Loeb}}{{K{\i}z{\i}ltan} et~al.}{2017}]{kiziltan17}
{K{\i}z{\i}ltan} B.,  {Baumgardt} H.,   {Loeb} A.,  2017, \mn@doi [\nat]
  {10.1038/nature21361}, \href
  {https://ui.adsabs.harvard.edu/abs/2017Natur.542..203K} {542, 203}

\bibitem[\protect\citeauthoryear{{Kulcs{\'a}r}, {Massioni}, {Sivo}  \&
  {Raynaud}}{{Kulcs{\'a}r} et~al.}{2012}]{kulcsar2012}
{Kulcs{\'a}r} C.,  {Massioni} P.,  {Sivo} G.,   {Raynaud} H.-F.~G.,  2012, in
  {Ellerbroek} B.~L.,  {Marchetti} E.,   {V{\'e}ran} J.-P.,  eds,  Society of
  Photo-Optical Instrumentation Engineers (SPIE) Conference Series Vol. 8447,
  Adaptive Optics Systems III. p. 84470Z, \mn@doi{10.1117/12.926050}

\bibitem[\protect\citeauthoryear{{Larkin} et~al.,}{{Larkin}
  et~al.}{2010}]{iris}
{Larkin} J.~E.,  et~al., 2010, in {McLean} I.~S.,  {Ramsay} S.~K.,   {Takami}
  H.,  eds,  Society of Photo-Optical Instrumentation Engineers (SPIE)
  Conference Series Vol. 7735, Ground-based and Airborne Instrumentation for
  Astronomy III. p. 773529 (\mn@eprint {arXiv} {1007.1973}),
  \mn@doi{10.1117/12.856305}

\bibitem[\protect\citeauthoryear{{Lenzen} et~al.,}{{Lenzen}
  et~al.}{2003}]{naco1}
{Lenzen} R.,  et~al., 2003, in {Iye} M.,  {Moorwood} A. F.~M.,  eds,  Society
  of Photo-Optical Instrumentation Engineers (SPIE) Conference Series Vol.
  4841, Instrument Design and Performance for Optical/Infrared Ground-based
  Telescopes. pp 944--952, \mn@doi{10.1117/12.460044}

\bibitem[\protect\citeauthoryear{{Leschinski} et~al.,}{{Leschinski}
  et~al.}{2016}]{simcado}
{Leschinski} K.,  et~al., 2016, in \procspie. p. 991124 (\mn@eprint {arXiv}
  {1609.01480}), \mn@doi{10.1117/12.2232483}

\bibitem[\protect\citeauthoryear{{Libralato} et~al.,}{{Libralato}
  et~al.}{2018}]{libralato18}
{Libralato} M.,  et~al., 2018, \mn@doi [\apj] {10.3847/1538-4357/aac6c0}, \href
  {https://ui.adsabs.harvard.edu/abs/2018ApJ...861...99L} {861, 99}

\bibitem[\protect\citeauthoryear{{Lu}, {Neichel}, {Anderson}, {Sinukoff},
  {Hosek}, {Ghez}  \& {Rigaut}}{{Lu} et~al.}{2014}]{lu14}
{Lu} J.~R.,  {Neichel} B.,  {Anderson} J.,  {Sinukoff} E.,  {Hosek} M.~W.,
  {Ghez} A.~M.,   {Rigaut} F.,  2014, in \procspie. p. 91480B,
  \mn@doi{10.1117/12.2057241}

\bibitem[\protect\citeauthoryear{{Marchetti} et~al.,}{{Marchetti}
  et~al.}{2003}]{madmcao}
{Marchetti} E.,  et~al., 2003, in {Wizinowich} P.~L.,  {Bonaccini} D.,  eds,
  Society of Photo-Optical Instrumentation Engineers (SPIE) Conference Series
  Vol. 4839, \procspie. pp 317--328, \mn@doi{10.1117/12.458859}

\bibitem[\protect\citeauthoryear{{Marchetti} et~al.,}{{Marchetti}
  et~al.}{2007}]{madonsky}
{Marchetti} E.,  et~al., 2007, The Messenger, \href
  {https://ui.adsabs.harvard.edu/abs/2007Msngr.129....8M} {129, 8}

\bibitem[\protect\citeauthoryear{{Massari} et~al.,}{{Massari}
  et~al.}{2015}]{massari2015}
{Massari} D.,  et~al., 2015, \mn@doi [\apj] {10.1088/0004-637X/810/1/69}, \href
  {https://ui.adsabs.harvard.edu/abs/2015ApJ...810...69M} {810, 69}

\bibitem[\protect\citeauthoryear{{Massari} et~al.,}{{Massari}
  et~al.}{2016a}]{massari16a}
{Massari} D.,  et~al., 2016a, \mn@doi [\aap] {10.1051/0004-6361/201527686},
  \href {https://ui.adsabs.harvard.edu/abs/2016A&A...586A..51M} {586, A51}

\bibitem[\protect\citeauthoryear{{Massari} et~al.,}{{Massari}
  et~al.}{2016b}]{massari16b}
{Massari} D.,  et~al., 2016b, \mn@doi [\aap] {10.1051/0004-6361/201629336},
  \href {https://ui.adsabs.harvard.edu/abs/2016A&A...595L...2M} {595, L2}

\bibitem[\protect\citeauthoryear{{Massari}, {Marasco}, {Beltramo-Martin},
  {Milli}, {Fiorentino}, {Tolstoy}  \& {Kerber}}{{Massari}
  et~al.}{2020}]{massari2020}
{Massari} D.,  {Marasco} A.,  {Beltramo-Martin} O.,  {Milli} J.,  {Fiorentino}
  G.,  {Tolstoy} E.,   {Kerber} F.,  2020, \mn@doi [\aap]
  {10.1051/0004-6361/201937359}, \href
  {https://ui.adsabs.harvard.edu/abs/2020A&A...634L...5M} {634, L5}

\bibitem[\protect\citeauthoryear{{McDermid} et~al.,}{{McDermid}
  et~al.}{2020}]{mavisscience}
{McDermid} R.~M.,  et~al., 2020, arXiv e-prints, \href
  {https://ui.adsabs.harvard.edu/abs/2020arXiv200909242M} {p. arXiv:2009.09242}

\bibitem[\protect\citeauthoryear{{McLaughlin}, {Anderson}, {Meylan},
  {Gebhardt}, {Pryor}, {Minniti}  \& {Phinney}}{{McLaughlin}
  et~al.}{2006}]{mclaughlin06}
{McLaughlin} D.~E.,  {Anderson} J.,  {Meylan} G.,  {Gebhardt} K.,  {Pryor} C.,
  {Minniti} D.,   {Phinney} S.,  2006, \mn@doi [\apjs] {10.1086/505692}, \href
  {https://ui.adsabs.harvard.edu/abs/2006ApJS..166..249M} {166, 249}

\bibitem[\protect\citeauthoryear{{Meyer}, {K{\"u}rster}, {Arcidiacono},
  {Ragazzoni}  \& {Rix}}{{Meyer} et~al.}{2011a}]{madastrometry}
{Meyer} E.,  {K{\"u}rster} M.,  {Arcidiacono} C.,  {Ragazzoni} R.,   {Rix}
  H.~W.,  2011a, \mn@doi [\aap] {10.1051/0004-6361/201016053}, \href
  {https://ui.adsabs.harvard.edu/abs/2011A&A...532A..16M} {532, A16}

\bibitem[\protect\citeauthoryear{{Meyer}, {K{\"u}rster}, {Arcidiacono},
  {Ragazzoni}  \& {Rix}}{{Meyer} et~al.}{2011b}]{madastrom}
{Meyer} E.,  {K{\"u}rster} M.,  {Arcidiacono} C.,  {Ragazzoni} R.,   {Rix}
  H.~W.,  2011b, \mn@doi [\aap] {10.1051/0004-6361/201016053}, \href
  {https://ui.adsabs.harvard.edu/abs/2011A&A...532A..16M} {532, A16}

\bibitem[\protect\citeauthoryear{{Miller} \& {Colbert}}{{Miller} \&
  {Colbert}}{2004}]{miller04}
{Miller} M.~C.,  {Colbert} E.~J.~M.,  2004, \mn@doi [International Journal of
  Modern Physics D] {10.1142/S0218271804004426}, \href
  {https://ui.adsabs.harvard.edu/abs/2004IJMPD..13....1M} {13, 1}

\bibitem[\protect\citeauthoryear{{Milone}, {Bedin}, {Piotto}  \&
  {Anderson}}{{Milone} et~al.}{2009}]{milone09}
{Milone} A.~P.,  {Bedin} L.~R.,  {Piotto} G.,   {Anderson} J.,  2009, \mn@doi
  [\aap] {10.1051/0004-6361/200810870}, \href
  {https://ui.adsabs.harvard.edu/abs/2009A&A...497..755M} {497, 755}

\bibitem[\protect\citeauthoryear{{Milone} et~al.,}{{Milone}
  et~al.}{2017}]{milone17}
{Milone} A.~P.,  et~al., 2017, \mn@doi [\mnras] {10.1093/mnras/stx836}, \href
  {https://ui.adsabs.harvard.edu/abs/2017MNRAS.469..800M} {469, 800}

\bibitem[\protect\citeauthoryear{{Monty} et~al.,}{{Monty}
  et~al.}{2018}]{monty2018}
{Monty} S.,  et~al., 2018, \mn@doi [\apj] {10.3847/1538-4357/aadb43}, \href
  {https://ui.adsabs.harvard.edu/abs/2018ApJ...865..160M} {865, 160}

\bibitem[\protect\citeauthoryear{{Neichel}, {Fusco}  \& {Conan}}{{Neichel}
  et~al.}{2008}]{neichel2008}
{Neichel} B.,  {Fusco} T.,   {Conan} J.-M.,  2008, \mn@doi [Journal of the
  Optical Society of America A] {10.1364/JOSAA.26.000219}, \href
  {https://ui.adsabs.harvard.edu/abs/2008JOSAA..26..219N} {26, 219}

\bibitem[\protect\citeauthoryear{{Neichel} et~al.,}{{Neichel}
  et~al.}{2014a}]{gems2}
{Neichel} B.,  et~al., 2014a, \mn@doi [\mnras] {10.1093/mnras/stu403}, \href
  {https://ui.adsabs.harvard.edu/abs/2014MNRAS.440.1002N} {440, 1002}

\bibitem[\protect\citeauthoryear{{Neichel}, {Lu}, {Rigaut}, {Ammons},
  {Carrasco}  \& {Lassalle}}{{Neichel} et~al.}{2014b}]{neichel14}
{Neichel} B.,  {Lu} J.~R.,  {Rigaut} F.,  {Ammons} S.~M.,  {Carrasco} E.~R.,
  {Lassalle} E.,  2014b, \mn@doi [\mnras] {10.1093/mnras/stu1766}, \href
  {https://ui.adsabs.harvard.edu/abs/2014MNRAS.445..500N} {445, 500}

\bibitem[\protect\citeauthoryear{{Neichel} et~al.,}{{Neichel}
  et~al.}{2021}]{neichel2021}
{Neichel} B.,  et~al., 2021, in Society of Photo-Optical Instrumentation
  Engineers (SPIE) Conference Series. p. 114482T (\mn@eprint {arXiv}
  {2101.06486}), \mn@doi{10.1117/12.2561533}

\bibitem[\protect\citeauthoryear{Noll}{Noll}{1976}]{noll76}
Noll R.~J.,  1976, \mn@doi [J. Opt. Soc. Am.] {10.1364/JOSA.66.000207}, 66, 207

\bibitem[\protect\citeauthoryear{{Noyola} \& {Baumgardt}}{{Noyola} \&
  {Baumgardt}}{2011}]{noyola2011}
{Noyola} E.,  {Baumgardt} H.,  2011, \mn@doi [\apj]
  {10.1088/0004-637X/743/1/52}, \href
  {https://ui.adsabs.harvard.edu/abs/2011ApJ...743...52N} {743, 52}

\bibitem[\protect\citeauthoryear{{Olivier}, {Max}, {Gavel}  \&
  {Brase}}{{Olivier} et~al.}{1993}]{scot93}
{Olivier} S.~S.,  {Max} C.~E.,  {Gavel} D.~T.,   {Brase} J.~M.,  1993, \mn@doi
  [\apj] {10.1086/172525}, \href
  {https://ui.adsabs.harvard.edu/abs/1993ApJ...407..428O} {407, 428}

\bibitem[\protect\citeauthoryear{{Patat}}{{Patat}}{2004}]{patat04}
{Patat} F.,  2004, The Messenger, \href
  {https://ui.adsabs.harvard.edu/abs/2004Msngr.115...18P} {115, 18}

\bibitem[\protect\citeauthoryear{{Patti} \& {Fiorentino}}{{Patti} \&
  {Fiorentino}}{2019}]{patti2019}
{Patti} M.,  {Fiorentino} G.,  2019, \mn@doi [\mnras] {10.1093/mnras/stz596},
  \href {https://ui.adsabs.harvard.edu/abs/2019MNRAS.485.3470P} {485, 3470}

\bibitem[\protect\citeauthoryear{{Piotto} et~al.,}{{Piotto}
  et~al.}{2015}]{piotto15}
{Piotto} G.,  et~al., 2015, \mn@doi [\aj] {10.1088/0004-6256/149/3/91}, \href
  {https://ui.adsabs.harvard.edu/abs/2015AJ....149...91P} {149, 91}

\bibitem[\protect\citeauthoryear{{Portegies Zwart}, {Baumgardt}, {Hut},
  {Makino}  \& {McMillan}}{{Portegies Zwart} et~al.}{2004}]{portegies04}
{Portegies Zwart} S.~F.,  {Baumgardt} H.,  {Hut} P.,  {Makino} J.,   {McMillan}
  S. L.~W.,  2004, \mn@doi [\nat] {10.1038/nature02448}, \href
  {https://ui.adsabs.harvard.edu/abs/2004Natur.428..724P} {428, 724}

\bibitem[\protect\citeauthoryear{{Raso} et~al.,}{{Raso}
  et~al.}{2020}]{raso2020}
{Raso} S.,  et~al., 2020, \mn@doi [\apj] {10.3847/1538-4357/ab8ae7}, \href
  {https://ui.adsabs.harvard.edu/abs/2020ApJ...895...15R} {895, 15}

\bibitem[\protect\citeauthoryear{{Rees}}{{Rees}}{1978}]{rees78}
{Rees} M.~J.,  1978, The Observatory, \href
  {https://ui.adsabs.harvard.edu/abs/1978Obs....98..210R} {98, 210}

\bibitem[\protect\citeauthoryear{{Rees}}{{Rees}}{1984}]{rees84}
{Rees} M.~J.,  1984, \mn@doi [\araa] {10.1146/annurev.aa.22.090184.002351},
  \href {https://ui.adsabs.harvard.edu/abs/1984ARA&A..22..471R} {22, 471}

\bibitem[\protect\citeauthoryear{{Riechert}, {Garrel}, {Pott}, {Sivo}  \&
  {Marin}}{{Riechert} et~al.}{2018}]{gemsmask}
{Riechert} H.,  {Garrel} V.,  {Pott} J.-U.,  {Sivo} G.,   {Marin} E.,  2018, in
  {Evans} C.~J.,  {Simard} L.,   {Takami} H.,  eds,  Society of Photo-Optical
  Instrumentation Engineers (SPIE) Conference Series Vol. 10702, Ground-based
  and Airborne Instrumentation for Astronomy VII. p. 1070232,
  \mn@doi{10.1117/12.2313460}

\bibitem[\protect\citeauthoryear{{Rigaut} \& {Neichel}}{{Rigaut} \&
  {Neichel}}{2018}]{rigaut2018}
{Rigaut} F.,  {Neichel} B.,  2018, \mn@doi [\araa]
  {10.1146/annurev-astro-091916-055320}, \href
  {https://ui.adsabs.harvard.edu/abs/2018ARA&A..56..277R} {56, 277}

\bibitem[\protect\citeauthoryear{{Rigaut}, {Veran}  \& {Lai}}{{Rigaut}
  et~al.}{1998}]{rigaut1998}
{Rigaut} F.~J.,  {Veran} J.-P.,   {Lai} O.,  1998, in {Bonaccini} D.,  {Tyson}
  R.~K.,  eds,  Society of Photo-Optical Instrumentation Engineers (SPIE)
  Conference Series Vol. 3353, Adaptive Optical System Technologies. pp
  1038--1048, \mn@doi{10.1117/12.321649}

\bibitem[\protect\citeauthoryear{{Rigaut} et~al.,}{{Rigaut}
  et~al.}{2014}]{gems1}
{Rigaut} F.,  et~al., 2014, \mn@doi [\mnras] {10.1093/mnras/stt2054}, \href
  {https://ui.adsabs.harvard.edu/abs/2014MNRAS.437.2361R} {437, 2361}

\bibitem[\protect\citeauthoryear{{Rigaut} et~al.,}{{Rigaut}
  et~al.}{2020}]{rigautmavis}
{Rigaut} F.,  et~al., 2020, in Society of Photo-Optical Instrumentation
  Engineers (SPIE) Conference Series. p. 114471R, \mn@doi{10.1117/12.2561886}

\bibitem[\protect\citeauthoryear{{Roddier}}{{Roddier}}{1981}]{roddier}
{Roddier} F.,  1981, \mn@doi [Progess in Optics]
  {10.1016/S0079-6638(08)70204-X}, \href
  {https://ui.adsabs.harvard.edu/abs/1981PrOpt..19..281R} {19, 281}

\bibitem[\protect\citeauthoryear{{Rodeghiero}, {Pott}, {Arcidiacono},
  {Massari}, {Gl{\"u}ck}, {Riechert}  \& {Gendron}}{{Rodeghiero}
  et~al.}{2018}]{micadodist}
{Rodeghiero} G.,  {Pott} J.~U.,  {Arcidiacono} C.,  {Massari} D.,  {Gl{\"u}ck}
  M.,  {Riechert} H.,   {Gendron} E.,  2018, \mn@doi [\mnras]
  {10.1093/mnras/sty1426}, \href
  {https://ui.adsabs.harvard.edu/abs/2018MNRAS.479.1974R} {479, 1974}

\bibitem[\protect\citeauthoryear{{Rodeghiero} et~al.,}{{Rodeghiero}
  et~al.}{2019}]{micadomask}
{Rodeghiero} G.,  et~al., 2019, \mn@doi [\pasp] {10.1088/1538-3873/ab0c40},
  \href {https://ui.adsabs.harvard.edu/abs/2019PASP..131e4503R} {131, 054503}

\bibitem[\protect\citeauthoryear{{Rousset} et~al.,}{{Rousset}
  et~al.}{2003a}]{vltao}
{Rousset} G.,  et~al., 2003a, in {Wizinowich} P.~L.,  {Bonaccini} D.,  eds,
  Society of Photo-Optical Instrumentation Engineers (SPIE) Conference Series
  Vol. 4839, Adaptive Optical System Technologies II. pp 140--149,
  \mn@doi{10.1117/12.459332}

\bibitem[\protect\citeauthoryear{{Rousset} et~al.,}{{Rousset}
  et~al.}{2003b}]{naco2}
{Rousset} G.,  et~al., 2003b, in {Wizinowich} P.~L.,  {Bonaccini} D.,  eds,
  Society of Photo-Optical Instrumentation Engineers (SPIE) Conference Series
  Vol. 4839, Adaptive Optical System Technologies II. pp 140--149,
  \mn@doi{10.1117/12.459332}

\bibitem[\protect\citeauthoryear{{Sakurai}, {Yoshida}, {Fujii}  \&
  {Hirano}}{{Sakurai} et~al.}{2017}]{sakurai17}
{Sakurai} Y.,  {Yoshida} N.,  {Fujii} M.~S.,   {Hirano} S.,  2017, \mn@doi
  [\mnras] {10.1093/mnras/stx2044}, \href
  {https://ui.adsabs.harvard.edu/abs/2017MNRAS.472.1677S} {472, 1677}

\bibitem[\protect\citeauthoryear{{Saracino} et~al.,}{{Saracino}
  et~al.}{2015}]{saracino15}
{Saracino} S.,  et~al., 2015, \mn@doi [\apj] {10.1088/0004-637X/806/2/152},
  \href {https://ui.adsabs.harvard.edu/abs/2015ApJ...806..152S} {806, 152}

\bibitem[\protect\citeauthoryear{{Saracino} et~al.,}{{Saracino}
  et~al.}{2016}]{saracino16}
{Saracino} S.,  et~al., 2016, \mn@doi [\apj] {10.3847/0004-637X/832/1/48},
  \href {https://ui.adsabs.harvard.edu/abs/2016ApJ...832...48S} {832, 48}

\bibitem[\protect\citeauthoryear{{Saracino} et~al.,}{{Saracino}
  et~al.}{2019}]{saracino19}
{Saracino} S.,  et~al., 2019, \mn@doi [\apj] {10.3847/1538-4357/ab07c4}, \href
  {https://ui.adsabs.harvard.edu/abs/2019ApJ...874...86S} {874, 86}

\bibitem[\protect\citeauthoryear{{Sarajedini} et~al.,}{{Sarajedini}
  et~al.}{2007}]{sarajedini07}
{Sarajedini} A.,  et~al., 2007, \mn@doi [\aj] {10.1086/511979}, \href
  {https://ui.adsabs.harvard.edu/abs/2007AJ....133.1658S} {133, 1658}

\bibitem[\protect\citeauthoryear{{Sch{\"o}ck}, {Do}, {Ellerbroek}, {Herriot},
  {Meyer}, {Suzuki}, {Wang}  \& {Yelda}}{{Sch{\"o}ck}
  et~al.}{2013}]{tmtastrombud1}
{Sch{\"o}ck} M.,  {Do} T.,  {Ellerbroek} B.,  {Herriot} G.,  {Meyer} L.,
  {Suzuki} R.,  {Wang} L.,   {Yelda} S.,  2013, in {Esposito} S.,  {Fini} L.,
  eds, Proceedings of the Third AO4ELT Conference. p.~77,
  \mn@doi{10.12839/AO4ELT3.13356}

\bibitem[\protect\citeauthoryear{{Sch{\"o}ck} et~al.,}{{Sch{\"o}ck}
  et~al.}{2016}]{tmtastrombud2}
{Sch{\"o}ck} M.,  et~al., 2016, in {Evans} C.~J.,  {Simard} L.,   {Takami} H.,
  eds,  Society of Photo-Optical Instrumentation Engineers (SPIE) Conference
  Series Vol. 9908, Ground-based and Airborne Instrumentation for Astronomy VI.
  p. 9908AD (\mn@eprint {arXiv} {1608.01693}), \mn@doi{10.1117/12.2231914}

\bibitem[\protect\citeauthoryear{{Service}, {Chun}, {Lu}, {Abdurrahman}, {Lai},
  {Fohring}  \& {Baranec}}{{Service} et~al.}{2018}]{tmtmask1}
{Service} M.,  {Chun} M.,  {Lu} J.,  {Abdurrahman} F.,  {Lai} O.,  {Fohring}
  D.,   {Baranec} C.,  2018, in Adaptive Optics Systems VI. p. 107034S,
  \mn@doi{10.1117/12.2313769}

\bibitem[\protect\citeauthoryear{{Service}, {Lu}, {Chun}, {Suzuki}, {Schoeck},
  {Atwood}, {Andersen}  \& {Herriot}}{{Service} et~al.}{2019}]{tmtmask2}
{Service} M.,  {Lu} J.~R.,  {Chun} M.,  {Suzuki} R.,  {Schoeck} M.,  {Atwood}
  J.,  {Andersen} D.,   {Herriot} G.,  2019, \mn@doi [Journal of Astronomical
  Telescopes, Instruments, and Systems] {10.1117/1.JATIS.5.3.039005}, \href
  {https://ui.adsabs.harvard.edu/abs/2019JATIS...5c9005S} {5, 039005}

\bibitem[\protect\citeauthoryear{{Stetson}}{{Stetson}}{1987}]{daophot1}
{Stetson} P.~B.,  1987, \mn@doi [\pasp] {10.1086/131977}, \href
  {https://ui.adsabs.harvard.edu/abs/1987PASP...99..191S} {99, 191}

\bibitem[\protect\citeauthoryear{{Stetson}}{{Stetson}}{1994}]{daophot2}
{Stetson} P.~B.,  1994, \mn@doi [\pasp] {10.1086/133378}, \href
  {https://ui.adsabs.harvard.edu/abs/1994PASP..106..250S} {106, 250}

\bibitem[\protect\citeauthoryear{{Str{\"o}bele} et~al.,}{{Str{\"o}bele}
  et~al.}{2012}]{galacsi2}
{Str{\"o}bele} S.,  et~al., 2012, in {Ellerbroek} B.~L.,  {Marchetti} E.,
  {V{\'e}ran} J.-P.,  eds,  Society of Photo-Optical Instrumentation Engineers
  (SPIE) Conference Series Vol. 8447, Adaptive Optics Systems III. p. 844737,
  \mn@doi{10.1117/12.926110}

\bibitem[\protect\citeauthoryear{{Suzuki}, {Harakawa}, {Sch\"{o}eck}  \&
  {Ellerbroek}}{{Suzuki} et~al.}{2017}]{tmta04elt5}
{Suzuki} R.,  {Harakawa} H.,  {Sch\"{o}eck} M.,   {Ellerbroek} B.,  2017, in
  Proceedings of the 5th Adaptive Optics for Extremely Large Telescopes
  Conference, Tenerife, Canary Islands, Spain, June 25-30.. ,
  \mn@doi{10.26698/AO4ELT5.0071}

\bibitem[\protect\citeauthoryear{{Thomas}, {Gavel}, {Adkins}  \&
  {Kibrick}}{{Thomas} et~al.}{2008}]{sandrine08}
{Thomas} S.~J.,  {Gavel} D.,  {Adkins} S.,   {Kibrick} B.,  2008, in {Hubin}
  N.,  {Max} C.~E.,   {Wizinowich} P.~L.,  eds,  Society of Photo-Optical
  Instrumentation Engineers (SPIE) Conference Series Vol. 7015, Adaptive Optics
  Systems. p. 70155L, \mn@doi{10.1117/12.789661}

\bibitem[\protect\citeauthoryear{{Toyozumi} \& {Ashley}}{{Toyozumi} \&
  {Ashley}}{2005}]{chargedif}
{Toyozumi} H.,  {Ashley} M. C.~B.,  2005, \mn@doi [\pasp] {10.1071/AS05013},
  22, 257–266

\bibitem[\protect\citeauthoryear{{Trippe}, {Davies}, {Eisenhauer}, {F{\"o}rster
  Schreiber}, {Fritz}  \& {Genzel}}{{Trippe} et~al.}{2010}]{trippe2010}
{Trippe} S.,  {Davies} R.,  {Eisenhauer} F.,  {F{\"o}rster Schreiber} N.~M.,
  {Fritz} T.~K.,   {Genzel} R.,  2010, \mn@doi [\mnras]
  {10.1111/j.1365-2966.2009.15940.x}, \href
  {https://ui.adsabs.harvard.edu/abs/2010MNRAS.402.1126T} {402, 1126}

\bibitem[\protect\citeauthoryear{{Turri}, {McConnachie}, {Stetson},
  {Fiorentino}, {Andersen}, {V{\'e}ran}  \& {Bono}}{{Turri}
  et~al.}{2015}]{turri15}
{Turri} P.,  {McConnachie} A.~W.,  {Stetson} P.~B.,  {Fiorentino} G.,
  {Andersen} D.~R.,  {V{\'e}ran} J.~P.,   {Bono} G.,  2015, \mn@doi [\apjl]
  {10.1088/2041-8205/811/2/L15}, \href
  {https://ui.adsabs.harvard.edu/abs/2015ApJ...811L..15T} {811, L15}

\bibitem[\protect\citeauthoryear{{Usher}, {Kamann}, {Gieles},
  {H{\'e}nault-Brunet}, {Dalessandro}, {Balbinot}  \& {Sollima}}{{Usher}
  et~al.}{2021}]{usher2021}
{Usher} C.,  {Kamann} S.,  {Gieles} M.,  {H{\'e}nault-Brunet} V.,
  {Dalessandro} E.,  {Balbinot} E.,   {Sollima} A.,  2021, \mn@doi [\mnras]
  {10.1093/mnras/stab565}, \href
  {https://ui.adsabs.harvard.edu/abs/2021MNRAS.503.1680U} {503, 1680}

\bibitem[\protect\citeauthoryear{{Veran}, {Rigaut}, {Maitre}  \&
  {Rouan}}{{Veran} et~al.}{1997}]{veran1997}
{Veran} J.~P.,  {Rigaut} F.,  {Maitre} H.,   {Rouan} D.,  1997, \mn@doi
  [Journal of the Optical Society of America] {10.1364/JOSAA.14.003057}, \href
  {https://ui.adsabs.harvard.edu/abs/1997OSAJ...14.3057V} {14, 3057}

\bibitem[\protect\citeauthoryear{Virtanen et~al.,}{Virtanen
  et~al.}{2020}]{scipy}
Virtanen P.,  et~al., 2020, \mn@doi [Nature Methods]
  {10.1038/s41592-019-0686-2}, \href {https://rdcu.be/b08Wh} {17, 261}

\bibitem[\protect\citeauthoryear{{Volonteri}}{{Volonteri}}{2010}]{volonteri10}
{Volonteri} M.,  2010, \mn@doi [\aapr] {10.1007/s00159-010-0029-x}, \href
  {https://ui.adsabs.harvard.edu/abs/2010A&ARv..18..279V} {18, 279}

\bibitem[\protect\citeauthoryear{{Watkins}, {van der Marel}, {Bellini}  \&
  {Anderson}}{{Watkins} et~al.}{2015}]{hstpromo2}
{Watkins} L.~L.,  {van der Marel} R.~P.,  {Bellini} A.,   {Anderson} J.,  2015,
  \mn@doi [\apj] {10.1088/0004-637X/803/1/29}, \href
  {https://ui.adsabs.harvard.edu/abs/2015ApJ...803...29W} {803, 29}

\bibitem[\protect\citeauthoryear{{Wizinowich} et~al.,}{{Wizinowich}
  et~al.}{2006}]{keckao}
{Wizinowich} P.~L.,  et~al., 2006, \mn@doi [\pasp] {10.1086/499290}, \href
  {https://ui.adsabs.harvard.edu/abs/2006PASP..118..297W} {118, 297}

\bibitem[\protect\citeauthoryear{{de Rijcke}, {Buyle}  \& {Dejonghe}}{{de
  Rijcke} et~al.}{2006}]{derijcke06}
{de Rijcke} S.,  {Buyle} P.,   {Dejonghe} H.,  2006, \mn@doi [\mnras]
  {10.1111/j.1745-3933.2006.00153.x}, \href
  {https://ui.adsabs.harvard.edu/abs/2006MNRAS.368L..43D} {368, L43}

\bibitem[\protect\citeauthoryear{{van den Born} \& {Jellema}}{{van den Born} \&
  {Jellema}}{2020}]{bornadc}
{van den Born} J.~A.,  {Jellema} W.,  2020, \mn@doi [\mnras]
  {10.1093/mnras/staa1870}, \href
  {https://ui.adsabs.harvard.edu/abs/2020MNRAS.496.4266V} {496, 4266}

\makeatother
\end{thebibliography}








\bsp	
\label{lastpage}
\end{document}